\documentclass[twocolumn]{aastex63}

\usepackage{lipsum, babel}

\usepackage{placeins}

\usepackage{array}

\newcommand\tess{TESS}
\newcommand\gaia{\textit{Gaia}}

\newcommand{\UCI}{Department of Physics \& Astronomy, The University of California, Irvine, Irvine, CA 92697, USA}
\newcommand{\IAC}{Instituto de Astrof\'\i sica de Canarias, E-38205, La Laguna, Tenerife, Spain}

\newcolumntype{P}[1]{>{\centering\arraybackslash}p{#1}}

\submitjournal{ApJ}

\begin{document}

\title{The TESS-Keck Survey XVII: Precise Mass Measurements in a Young, High Multiplicity Transiting Planet System using Radial Velocities and Transit Timing Variations}

\author[0000-0001-7708-2364]{Corey Beard}
\altaffiliation{NASA FINESST Fellow}
\affiliation{\UCI}
%confirmed

\author[0000-0003-0149-9678]{Paul Robertson}
\affiliation{\UCI}
%confirmed

\author[0000-0002-8958-0683]{Fei Dai}
\altaffiliation{NASA Sagan Fellow}
\affiliation{Division of Geological and Planetary Sciences,
1200 E California Blvd, Pasadena, CA, 91125, USA}
\affiliation{Department of Astronomy, California Institute of Technology, Pasadena, CA 91125, USA}
%confirmed

%%%%%%%%%%%%%%%%%%%%%

\author[0000-0002-5034-9476]{Rae Holcomb}
\affil{\UCI}
%confirmed

\author[0000-0001-8342-7736]{Jack Lubin}
\affil{\UCI}
%confirmed

%%%%%%%%%%%%%%%%%%%%%%

\author[0000-0001-8898-8284]{Joseph M. Akana Murphy}
\altaffiliation{NSF Graduate Research Fellow}
\affiliation{Department of Astronomy and Astrophysics, University of California, Santa Cruz, CA 95064, USA}
%confirmed

\author[0000-0002-7030-9519]{Natalie M. Batalha}
\affiliation{Department of Astronomy and Astrophysics, University of California, Santa Cruz, CA 95064, USA}
%confirmed

\author[0000-0002-3199-2888]{Sarah Blunt}
\affiliation{Department of Astronomy, California Institute of Technology, Pasadena, CA 91125, USA}
%confirmed

\author{Ian Crossfield}
\affiliation{Department of Physics and Astronomy, University of Kansas, Lawrence, KS, USA}
%confirmed
	
\author{Courtney Dressing}
\affiliation{501 Campbell Hall, University of California at Berkeley, Berkeley, CA 94720, USA}

\author[0000-0003-3504-5316]{Benjamin Fulton}
\affiliation{NASA Exoplanet Science Institute/Caltech-IPAC, MC 314-6, 1200 E California Blvd, Pasadena, CA 91125, USA}
%confirmed

\author[0000-0001-8638-0320]{Andrew W. Howard}
\affiliation{California Institute of Technology, Pasadena, CA 91125, USA}
%confirmed

\author{Dan Huber}
\affiliation{Institute for Astronomy, University of Hawai'i, 2680 Woodlawn Drive, Honolulu, HI 96822 USA}
%confirmed

\author[0000-0002-0531-1073]{Howard Isaacson}
\affiliation{501 Campbell Hall, University of California at Berkeley, Berkeley, CA 94720, USA}
\affiliation{Centre for Astrophysics, University of Southern Queensland, Toowoomba, QLD, Australia}
%confirmed

\author[0000-0002-7084-0529]{Stephen R. Kane}
\affiliation{Department of Earth and Planetary Sciences, University of California, Riverside, CA 92521, USA}
%confirmed

\author[0000-0002-7031-7754]{Grzegorz Nowak}
\affiliation{Institute of Astronomy, Faculty of Physics, Astronomy and Informatics, Nicolaus Copernicus University, Grudzi\c{a}dzka 5, 87-100 Toru\'n, Poland}
\affiliation{\IAC}
\affiliation{Departamento de Astrof\'isica, Universidad de La Laguna, Av. Astrof\'isico Francisco S\'anchez, s/n, E-38206 La Laguna, Tenerife, Spain}
%confirmed

\author[0000-0003-0967-2893]{Erik A Petigura}
\affiliation{Department of Physics \& Astronomy, University of California Los Angeles, Los Angeles, CA 90095, USA}
%confirmed

\author[0000-0001-8127-5775]{Arpita Roy}
\affiliation{Space Telescope Science Institute, 3700 San Martin Drive, Baltimore, MD 21218, USA}
\affiliation{Department of Physics and Astronomy, Johns Hopkins University, 3400 N Charles St, Baltimore, MD 21218, USA}
%confirmed

\author[0000-0003-3856-3143]{Ryan A. Rubenzahl}
\altaffiliation{NSF Graduate Research Fellow}
\affiliation{Department of Astronomy, California Institute of Technology, Pasadena, CA 91125, USA}
%confirmed

\author[0000-0002-3725-3058]{Lauren M. Weiss}
\affiliation{Department of Physics and Astronomy, University of Notre Dame, Notre Dame, IN 46556, USA}
%confirmed

%%%%%%%%%%%%%%%%%%%%%%%%%%%

\author{Rafael Barrena}
\affil{\IAC}
\affil{Universidad de La Laguna, Departamento de Astrof\'\i sica, E-38206, La Laguna, Tenerife, Spain}
%confirmed

\author[0000-0003-0012-9093]{Aida Behmard}
\affiliation{Division of Geological and Planetary Sciences,
1200 E California Blvd, Pasadena, CA, 91125, USA}
%confirmed

\author[0000-0002-4480-310X]{Casey L. Brinkman}
\affiliation{Institute for Astronomy, University of Hawai'i, 2680 Woodlawn Drive, Honolulu, HI 96822 USA}
%confirmed

\author{Ilaria Carleo}
\affil{\IAC}
\affil{Universidad de La Laguna, Departamento de Astrof\'\i sica, E-38206, La Laguna, Tenerife, Spain}
%confirmed

\author[0000-0003-1125-2564]{Ashley Chontos}
\altaffiliation{Henry Norris Russell Fellow}
\affiliation{Department of Astrophysical Sciences, Princeton University, 4 Ivy Lane, Princeton, NJ, 08544, USA}
\affiliation{Institute for Astronomy, University of Hawai`i, 2680 Woodlawn Drive, Honolulu, HI 96822, USA}
%confirmed

\author[0000-0002-4297-5506]{Paul A.\ Dalba}
\altaffiliation{Heising-Simons 51 Pegasi b Postdoctoral Fellow}
\affiliation{Department of Astronomy and Astrophysics, University of California, Santa Cruz, CA 95064, USA}
\affiliation{SETI Institute, Carl Sagan Center, 339 Bernardo Ave, Suite 200, Mountain View, CA 94043, USA}
%confirmed

\author[0000-0002-3551-279X]{Tara Fetherolf}
\altaffiliation{UC Chancellor's Fellow}
\affiliation{Department of Earth and Planetary Sciences, University of California, Riverside, CA 92521, USA}
%confirmed

\author[0000-0002-8965-3969]{Steven Giacalone}
\affil{Department of Astronomy, University of California Berkeley, Berkeley, CA 94720, USA}
%confirmed

\author[0000-0002-0139-4756]{Michelle L. Hill}
\affiliation{Department of Earth and Planetary Sciences, University of California, Riverside, CA 92521, USA}
%confirmed

\author{Kiyoe Kawauchi}
\affiliation{Department of Physical Sciences, Ritsumeikan University, Kusatsu, Shiga 525-8577, Japan}
%confirmed

\author[0000-0002-0076-6239]{Judith Korth}
\affiliation{Lund Observatory, Division of Astrophysics, Department of Physics, Lund University, Box 43, 22100 Lund, Sweden}
%confirmed

\author[0000-0002-4671-2957]{Rafael Luque}
\affiliation{Department of Astronomy \& Astrophysics, University of Chicago, Chicago, IL 60637, USA}
%confirmed

\author[0000-0003-2562-9043]{Mason G. MacDougall}
\affiliation{Department of Physics \& Astronomy, University of California Los Angeles, Los Angeles, CA 90095, USA}
%confirmed

\author[0000-0002-7216-2135]{Andrew W. Mayo}
\affiliation{501 Campbell Hall, University of California at Berkeley, Berkeley, CA 94720, USA}
%confirmed

\author[0000-0003-4603-556X]{Teo Mo\v{c}nik}
\affiliation{Gemini Observatory/NSF's NOIRLab, 670 N. A'ohoku Place, Hilo, HI 96720, USA}
%confirmed

\author[0000-0002-4262-5661]{Giuseppe Morello}
\affiliation{Department of Space, Earth and Environment, Chalmers University of Technology, SE-412 96 Gothenburg, Sweden}
\affiliation{\IAC}
%confirmed

\author[0000-0001-9087-1245]{Felipe Murgas}
\affil{\IAC}
\affil{Universidad de La Laguna, Departamento de Astrof\'\i sica, E-38206, La Laguna, Tenerife, Spain}
%confirmed

\author[0000-0003-2066-8959]{Jaume Orell-Miquel}
\affiliation{\IAC}
%confirmed

\author{Enric Palle}
\affil{\IAC}
\affil{Universidad de La Laguna, Departamento de Astrof\'\i sica, E-38206, La Laguna, Tenerife, Spain}
%confirmed

\author[0000-0001-7047-8681]{Alex S. Polanski}
\affil{Department of Physics and Astronomy, University of Kansas, Lawrence, KS 66045, USA}
%confirmed

\author[0000-0002-7670-670X]{Malena Rice}
\altaffiliation{Heising-Simons 51 Pegasi b Postdoctoral Fellow}
\affiliation{Department of Astronomy, Yale University, New Haven, CT 06511, USA}
\affiliation{Department of Physics and Kavli Institute for Astrophysics and Space Research, Massachusetts Institute of Technology, Cambridge, MA 02139, USA}
%confirmed

\author[0000-0003-3623-7280]{Nicholas Scarsdale}
\affiliation{Department of Astronomy and Astrophysics, University of California, Santa Cruz, CA 95064, USA}
%confirmed

\author[0000-0003-0298-4667]{Dakotah Tyler}
\affiliation{Department of Physics \& Astronomy, University of California Los Angeles, Los Angeles, CA 90095, USA}
%confirmed

\author[0000-0002-4290-6826]{Judah Van Zandt}
\affil{Department of Physics \& Astronomy, University of California Los Angeles, Los Angeles, CA 90095, USA}
%confirmed

\correspondingauthor{Corey Beard}
\email{ccbeard@uci.edu}

\begin{abstract}

We present a radial velocity (RV) analysis of TOI-1136, a bright TESS system with six confirmed transiting planets, and a seventh single-transiting planet candidate. All planets in the system are amenable to transmission spectroscopy, making TOI-1136 one of the best targets for intra-system comparison of exoplanet atmospheres. TOI-1136 is young ($\sim$ 700 Myr), and the system exhibits transit timing variations (TTVs). The youth of the system contributes to high stellar variability on the order of 50 m s$^{-1}$, much larger than the likely RV amplitude of any of the transiting exoplanets. Utilizing 359 HIRES and APF RVs collected as a part of the TESS-Keck Survey (TKS), and 51 HARPS-N RVs, we experiment with a joint TTV-RV fit. With seven possible transiting planets, TTVs, more than 400 RVs, and a stellar activity model, we posit that we may be presenting the most complex mass recovery of an exoplanet system in the literature to date. By combining TTVs and RVs, we minimized GP overfitting and retrieved new masses for this system: (m$_{b-g}$ = 3.50$^{+0.8}_{-0.7}$, 6.32$^{+1.1}_{-1.3}$, 8.35$^{+1.8}_{-1.6}$, 6.07$^{+1.09}_{-1.01}$, 9.7$^{+3.9}_{-3.7}$, 5.6$^{+4.1}_{-3.2}$ M$_{\oplus}$). We are unable to significantly detect the mass of the seventh planet candidate in the RVs, but we are able to loosely constrain a possible orbital period near 80 days. Future TESS observations might confirm the existence of a seventh planet in the system, better constrain the masses and orbital properties of the known exoplanets, and generally shine light on this scientifically interesting system.

\end{abstract}


\section{Introduction} \label{sec:intro}

Among the most pressing scientific questions in the field of exoplanet science are those of planet formation and the subsequent evolution of planetary systems. Population studies are generally required to learn about such processes, as the involved astronomical timescales are far too long for direct observations. Multiple different processes might explain the formation of exoplanets in other systems: pebble accretion could form planets in situ \citep{lambrechts12, izidoro21}, or planets might form beyond the ice line and migrate inward \citep{goldreich80,dangelo03,raymond08,izidoro17}. The subsequent evolution of planets post formation, such as atmospheric sculpting from stellar flux \citep{lammer03} or the outgassing of volatiles \citep{rogers11}, likely also affects the observed planetary population. 

Challenges comparing different exoplanetary systems are compounded by the various natures of host stars: while it is possible to control for aspects such as age, stellar type, and metallicity when studying formation history, the probable dependence of system evolution on a variety of the host star's parameters makes wider study difficult. Consequently, multiple-planet systems are very attractive when studying planetary characteristics such as atmospheric evolution, as they all share a host star, allowing for the removal of degeneracies between different stellar parameters. Higher multiplicities are better, as they allow for a larger sample size that shares system parameters. 

The successful launch of JWST \citep{gardner06} places a particular emphasis on systems that are highly amenable to atmospheric characterization via transmission spectroscopy, as this is one of JWST's primary science goals. Early programs with JWST are already observing previously undetected exoplanet atmospheric features \citep{jwst23} and ruling out atmospheres in popular targets such as TRAPPIST-1 b \citep{greene23}.

Here we present a follow-up analysis of TOI-1136, a system with at least six transiting planets first characterized by \citet[][hereafter D23]{dai22}, and a candidate seventh. TOI-1136 is a young (700 $\pm$ 100 Myr), bright (V=9.5) G dwarf that has several planets that exhibit significant transit timing variations (TTVs), allowing for the precise characterization of most planet masses with photometry alone. The planets are in deep Laplace resonance (P$_{b}$ = 4.1727 d; P$_{c}$ = 6.2574 d; P$_{d}$ = 12.5199 d; P$_{e}$ = 18.801 d; P$_{f}$ = 26.321 d; P$_{g}$ = 39.545 d), suggesting a distinct formation history \citep[short scale type-I migration;][D23]{sinclair75}. TOI-1136 was observed by the Transiting Exoplanet Survey Satellite \citep[TESS;][]{ricker15} for six non-consecutive sectors. The relatively short baseline of TESS limits the precision with which we can constrain planetary masses using TTVs, especially for the longer-period outer planets. Furthermore, adding RVs in conjunction with TTVs can help prevent conflict between TTV-only and RV-only measured masses \citep{steffen16,mills17}.

The TESS-Keck Survey \citep[TKS;][]{chontos22} carried out extensive radial velocity (RV) observations of TOI-1136 as a part of its primary survey to measure the masses of 100 transiting planets. TKS is divided into a variety of science cases studying the radius gap first identified in \cite{fulton17} \citep{weiss21,brinkman23}, orbital dynamics \citep{rubenzahl21,macdougall21}, multi-planet systems \citep{lubin22,turtelboom22}, and atmospheres \citep[][]{scarsdale21,murphy23} to name a few. TOI-1136 fits almost every science case in TKS, and is consequently a very important system for the TKS team to understand.

We utilize over 400 RVs taken with the High Resolution Echelle Spectrometer \citep[HIRES; ][]{vogt94}, Levy Spectrometer on the robotic Automated Planet Finder \citep[APF;][]{vogt14} Telescope, and the High-Accuracy Radial velocity Planetary Searcher North \citep[HARPS-N;][]{cosentino12} spectrograph.  We combine these observations with TTVs and perform a detailed RV + TTV analysis of TOI-1136. 

The paper is organized as follows. A summary of our observations and data is given in \S\ref{sec:observations}. A brief description of the stellar parameters of TOI-1136 is given in \S\ref{sec:stellar}. A study of the candidate seventh planet is given in \S\ref{planet_h}. Our analysis is detailed in \S\ref{sec:analysis}. Finally, the results and their interpretation are placed into context in \S\ref{sec:discussion}, and the paper is summarized in \S\ref{sec:summary}.

\section{Observations}\label{sec:observations}

\subsection{TESS Photometry}

TOI-1136 was first observed by TESS during Sector 14 (18 July - 15 August 2019) and Sector 15 (15 August - 11 September 2019) of Cycle 2. TOI-1136 was later re-observed during Sectors 21 and 22 (21 January - 18 March 2020), and in two subsequent sectors (41: 23 July - 20 August 2021; 48: 28 January - 26 February 2022). The star was first declared a TESS object of interest \citep[TOI;][]{guerrero21} on 27 August 2019, and the science processing and operations center \citep[SPOC;][]{jenkins16} pipeline would eventually identify 4 candidate planets in the system. Additional community observers would later identify two more community TOIs (CTOIs), increasing the number of candidate planets in the system to 6. 

No additonal TESS photometry has been acquired since the system was studied in D23. Nonetheless, the TESS photometry is incorporated in multiple aspects of our analysis of the system.  We build on the individual transit times of the TOI-1136 planets determined in D23 by jointly modeling these transit times with RVs in \S \ref{sec:analysis}. We also analyze the TESS Presearch Data Cleaning Simple Aperture Flux \citep[PDCSAP;][]{jenkins16} photometry to measure the stellar rotation period, to fit a single transit to the candidate planet in \S \ref{sec:planet_h_period}, and to calculate FF$^{\prime}$ values utilized in \S \ref{sec:ffprime} by multiplying the PDCSAP Flux (F) by its first derivative \citep[F$^{\prime}$;][]{aigrain12}.

\subsection{RVs with Keck/HIRES}
\label{sec:hiresrvs}

Between 1 November 2019 and 16 July 2022, we obtained 155 high-resolution spectra of TOI-1136, resulting in 103 nightly binned RV observations, using the High-Resolution Echelle Spectrometer (HIRES, \citealt{vogt94}) located at Keck Observatory. Precise radial velocities were extracted using a warm iodine cell in the light path for wavelength calibration, as described in \cite{butler96}.  We extracted precise RVs from the echelle spectra using the California Planet Search (CPS) pipeline \citep{howard10}.

We typically achieved a signal-to-noise ratio of $\approx200$ at visible wavlengths for each spectrum by capping the HIRES built-in exposure meter at 250,000 counts.  resulting in a median nightly binned RV uncertainty of 1.75 m s$^{-1}$, and a median SNR of 214 for the wavelength order centered at 540 nm.

\subsection{HARPS-N RVs}
\label{sec:harpsnrvs}

We also utilized 51 RV observations of TOI-1136 obtained using the HARPS-N spectrograph at the Telescopio Nazionale Galileo, a 3.6-m telescope located in the Canary Islands, Spain under the observing programs, CAT19A\_162, ITP19\_1 and CAT21A\_119. Observations had a median exposure time of 1000 s and a median SNR of 74.6 at 550 nm.

HARPS-N RVs were reduced using the standard cross-correlation function mask method outlined in \cite{baranne96} and \cite{pepe02}. After reduction, HARPS-N RVs had a mean uncertainty of 2.63 m s$^{-1}$.

\subsection{RVs with the Automated Planet Finder}

Essential to characterizing the stellar activity were additional RV observations taken using the APF Telescope, located at Lick Observatory on Mount Hamilton, CA. The automated nature of the APF allowed for much more consistent, high cadence observations than were possible using HIRES or HARPS-N. The smaller aperture of APF, however, restricted us to lower SNR and correspondingly less precise observations. Between 1 November 2019 and 16 July 2022, we carried out 320 APF observations over the course of 256 observing nights. APF spectra are calibrated using an iodine cell and are extracted using a process very similar to that of HIRES RV extraction \citep{fulton15}.

A preliminary analysis of APF spectra motivated our choice of a minimum SNR threshold of 55, as spectra with lower SNR were subject to very large uncertainties. Our final collection of APF observations have a mean binned RV uncertainty of 4.92 m s$^{-1}$ and a mean SNR of 94.1 estimated across its full wavelength coverage, centered at 596 nm.

\section{Stellar Parameters}
\label{sec:stellar}

We utilize the stellar parameters of TOI-1136 as adopted in D23. D23 used \texttt{SpecMatchSyn} \citep{petigura17} on 3 iodine-free, high resolution HIRES spectra obtained as a part of TKS's observing program to derive $T_{\rm{eff}}$, $\log g$, and [Fe/H]. These results were then combined with Gaia parameters \citep{gaia16,gaiadr3} in the \textsf{Isoclassify} software package to obtain stellar mass, radius, and other relevant parameters for our models \citep{huber17,berger20_b}. The full list of stellar parameters is identical to those used in D23, and we refer readers to Table 1 in D23 for the full parameter list, though for convenience we note that the star is a G5 dwarf with M$_{*}$ = 1.022 $\pm$ 0.027 M$_{\odot}$ and R$_{*}$ = 0.968 $\pm$ 0.036 R$_{\odot}$. We detail our independent measurement of the system's stellar rotation in the next subsection.

\subsection{Stellar Rotation Period}
\label{sec:acf}
Identifying the frequency of stellar rotation is an important part of characterizing systems with spot modulation, as a quasiperiodic signal can mask known exoplanet signals \citep{lopezmorales16} or mimic real ones \citep{lubin21}. Young systems, like TOI-1136, are particularly susceptible to large activity signals that dwarf planetary signals \citep{cale21}.

\begin{figure}
    \centering
    \includegraphics[width=0.5\textwidth]{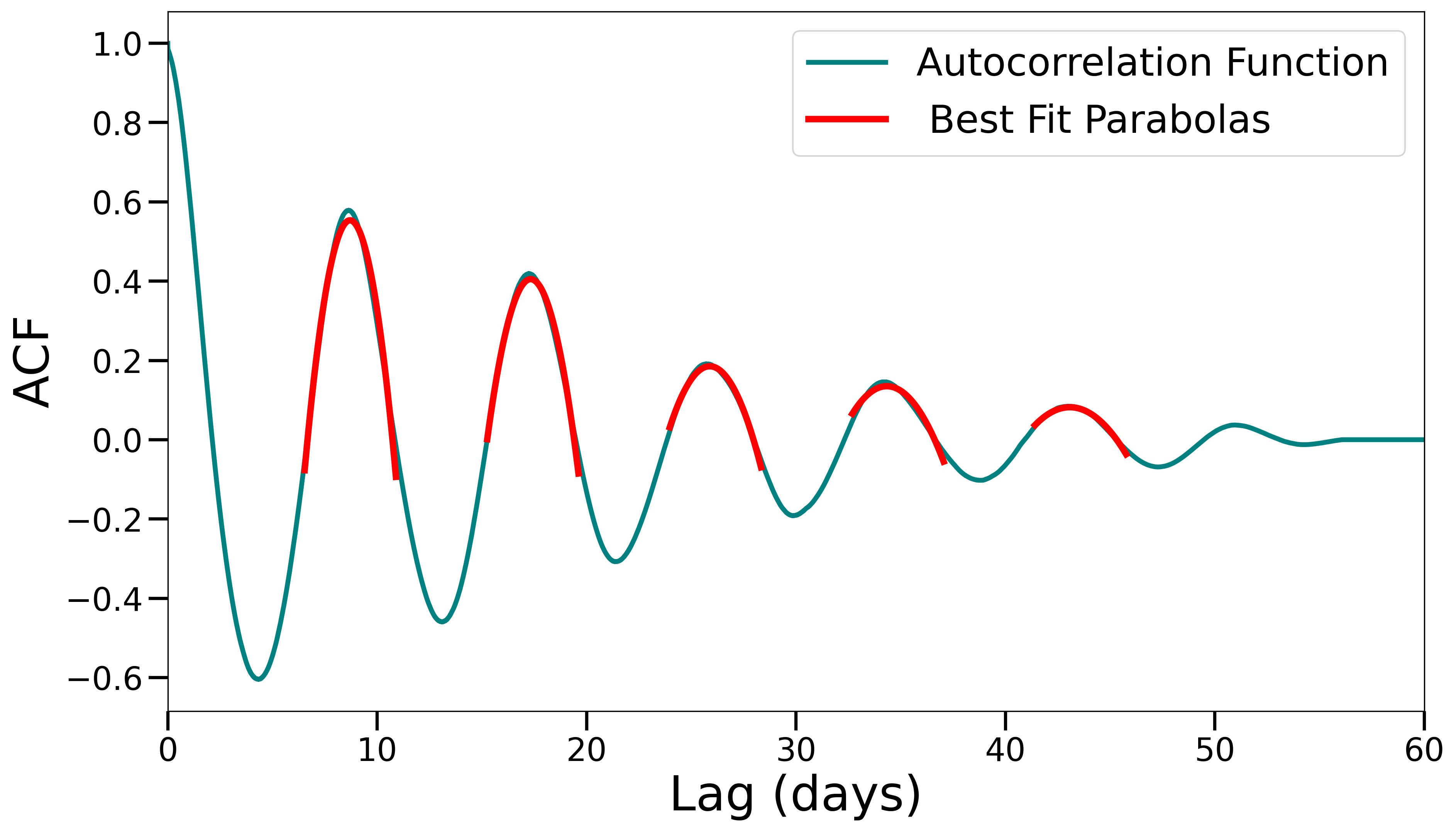}
    \caption{An autocorrelation function (ACF) of TOI-1136's TESS photometry. A clear frequency pattern with well-defined parabolas fit to the peaks of the ACF indicates a solid detection of the system's rotation period.}
    \label{fig:acf}
\end{figure}

D23 used a Lomb-Scargle periodogram \citep[LS;][]{lomb76, scargle82} to identify a rotation period of 8.7 $\pm$ 0.1 days for TOI-1136 based on TESS photometry. Because the quasi-periodic rotation signature often leads to significant peaks at harmonics of the true rotation period, we performed an independent analysis of the stellar rotation.  We utilized the new \textsf{SpinSpotter} software package to fit an autocorrelation function (ACF) to TESS photometry \citep{holcomb22}, which is more robust than LS for detecting accurate stellar rotation periods \citep{aigrain15}. We analyzed the ACF on all six Sectors of TESS data. We identified a rotation period of 8.42 $\pm$ 0.09 days using \textsf{SpinSpotter}, which is consistent with the previously identified rotation period (as opposed to a harmonic). The uncertainty is estimated by taking the standard deviation of the variations in parabola vertex locations from the expected position predicted by the found rotation period, which can underestimate uncertainties and likely contributes to the $> 1\,\sigma$ discrepancy with D23. However, \cite{holcomb22} suggest that detecting at least 5 peaks in the ACF (seen in Figure \ref{fig:acf}) is strong evidence that the rotation estimate is reliable.  

\section{A Seventh Planet?}
\label{planet_h}

D23 identified a single transit that was distinct from those corresponding to planets b-g as a possible seventh planet in the TOI-1136 system. D23 were unable to identify any additional transits of this candidate planet in the TESS photometry, and so the period remained unclear.

D23 did not include a detailed analysis of the transit, mainly noting that the estimated radius was around 2.5 R$_{\oplus}$ and that the transit duration suggested a possible $\sim$ 80 day orbital period, consistent with a 2:1 resonance with planet g. 

Without additional transits, it is difficult to conclude that the event is necessarily an exoplanet. False positive transit signatures were rare in NASA \textit{Kepler} photometry, but are more common in the TESS photometry due to a large pixel size \citep{sullivan15}. D23 ruled out visual, spectroscopic, co-moving, and astrometric companions, but this does not exclude every possible false positive scenario. For example, an unresolved background eclipsing binary could potentially create such a signature. False positives from background eclipsing binaries are extremely unlikely in the high-multiplicity planetary systems characterized by \textit{Kepler}\ \citep{lissauer12}, but the incidence of eclipsing binary FPs is likely higher for TESS planet candidates due to the larger pixel size.  Another possibility is that the transit-like event is a false alarm, i.e., an instrumental artifact or spurious event that is non-astrophysical.  To mitigate our uncertainty of the veracity of planet candidate seven, we tested both a seven-planet and a six-planet model in our full TTV + RV analyses, with constraints on the orbit of the seventh planet based on the RVs.

\subsection{Identifying the Period of the Candidate}
\label{sec:planet_h_period}

D23 estimate an orbital period near 80 days for the seventh planet candidate based on the transit duration. Such an estimate can be inaccurate, however, especially when factors such as eccentricity and impact parameter are also unknown. We explore other methods of estimating an orbital period for the single-transit candidate.

The period might be inferred from the RVs, as their quantity and cadence would be sufficient to find medium to longer-period planets with modest amplitudes in many planetary systems. A periodogram analysis is often fruitful when first trying to identify the orbital periods of planets in the RVs.

We first used a generalized Lomb-Scargle (GLS) periodogram \citep{zechmeister09} to identify significant periodicity in the RV data. Unfortunately, the GLS does not identify the periods of any of the known transiting planets, and it is unlikely that any of the high power periods correspond with the candidate planet. The highest-power periods are all close to, or aliases of, the known rotation period of TOI-1136. This is mainly caused by the prominent stellar activity in the system. Due to the quasi-periodic nature of the rotation signal, however, standard sinusoidal fits are an imperfect match, and one cannot easily subtract the highest power signal for investigation of lower amplitude signals. 

We next try the correlated noise model present in the Bayes Factor periodogram (BFP) with 1 moving average term introduced in \cite{feng17}, but this too proved insufficient to clearly detect the orbital period of any planet exterior to planet g. Most likely, the red-noise model used by the BFP, while consistently recovering true stellar variability signals, was not capable of detecting the relatively small amplitude of the planets in the system.

\begin{figure}
    \centering
    \includegraphics[width=0.5\textwidth]{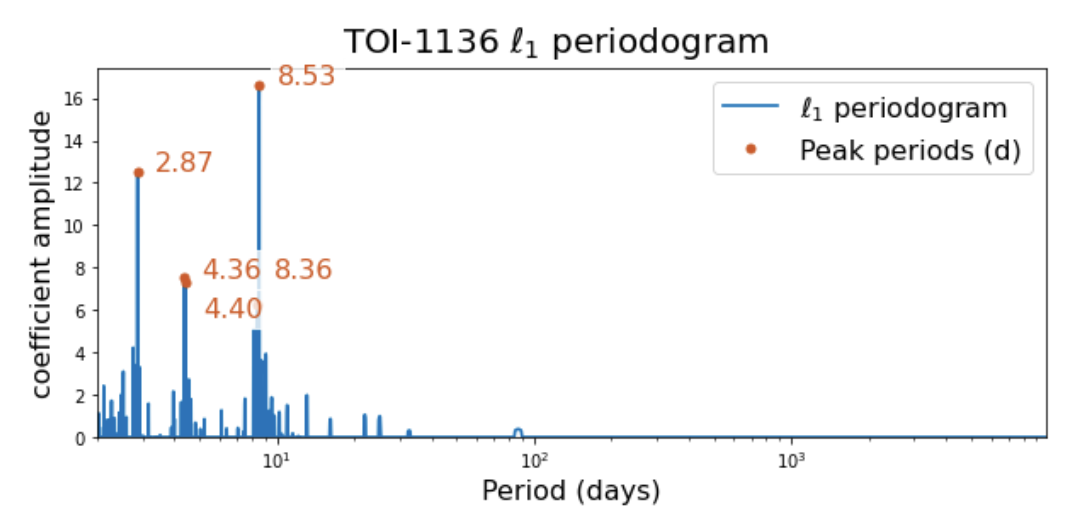}
    \caption{We computed an $l$1 periodogram of TOI-1136 RV data, determining the best white noise value for the noise model through cross-validation. Instrument offsets are fit by the compressed sensing model. Unlike other periodograms, multiple peaks can have significance. However, the rotation period (8.53 days), signals near the rotation period (8.36 days), and aliases of the rotation period (4.40, 4.36, 2.87 days) dominate the periodogram. Once again, planet periods are not significantly detected, and a more complicated model is required to remove the activity and uncover the planet signals.}
    \label{fig:l1periodogram}
\end{figure}

The $l$1 periodogram established in \cite{hara17} searches all frequencies simultaneously rather than sequentially, and might succeed where other frequency analysis attempts have failed. However, as is visible in Figure \ref{fig:l1periodogram}, the $l$1 periodogram once again only identifies the rotation period and its aliases, probably due to their much more significant amplitudes.

Another method we might use to predict the period of the candidate planet exploits the resonance of TOI-1136. For example, a similar method was used to predict the orbital period of TRAPPIST-1 h when only one transit of the planet was known \citep{luger17}. The idea is that trios of neighboring planets in compact, resonant systems tend to satisfy equation \ref{eqn:3body} for small-integer values of $p$, $q$.

\begin{equation}
    \label{eqn:3body}
    \centering
    pP_1^{-1} - (p+q)P_2^{-1} + qP_3^{-1} \approx 0
\end{equation}

Here $P_1$, $P_2$, and $P_3$ are the orbital periods of any three adjacent planets. We solved for $P_3$ for a variety of combinations of $p$ and $q$, ranging from one to three. Many of the predictions were implausible. Some combinations predicted orbital periods interior to planet g's $\sim 39$ day orbital period (e.g. $p=1$, $q=2$, $P_3=32.89$ days), which would have been seen in photometry, and are unlikely in such a compact, resonant system. Some predictions were close enough to planet g for stability concerns to make the period unlikely (e.g. $p=2$, $q=2$, $P_3=43.86$ days). Two period predictions stand out as plausible when using equation \ref{eqn:3body}: $p=2$, $q=1$, $P_3=131.47$ days and $p=3$, $q=2$, $P_3=65.71$ days. The first is somewhat close to 4:1 resonance with planet g at 156 days, and the second is close to 3:2 resonance at 58.5 days, or perhaps a 2:1 resonance at 79.1 days. Motivated by the $\sim$ 80 day period prediction from transit duration, we deem the 65.71 day period the more likely of the two. We proceed assuming the candidate planet is either in 3:2 or 2:1 resonance with planet g.

\subsection{Fitting the Transit of the Candidate Planet}
\label{h_radius}

We have not identified any additional transits of the candidate, though our RV analysis in section \ref{sec:ttv_rv} does shed additional light on the planet. In order to frame the candidate in context with the other planets in TOI-1136, we perform a single transit fit to formalize an estimate of the radius and other transit-related parameters.

We use the \texttt{exoplanet} software package \citep{Foreman-Mackey2021, exoplanet:zenodo} on the detrended photometry from D23, which is detrended using a simple 0.5 day cubic spline. We used only photometry within one day of the reported transit time in D23. 

\begin{figure}
    \centering
    \includegraphics[width=0.5\textwidth]{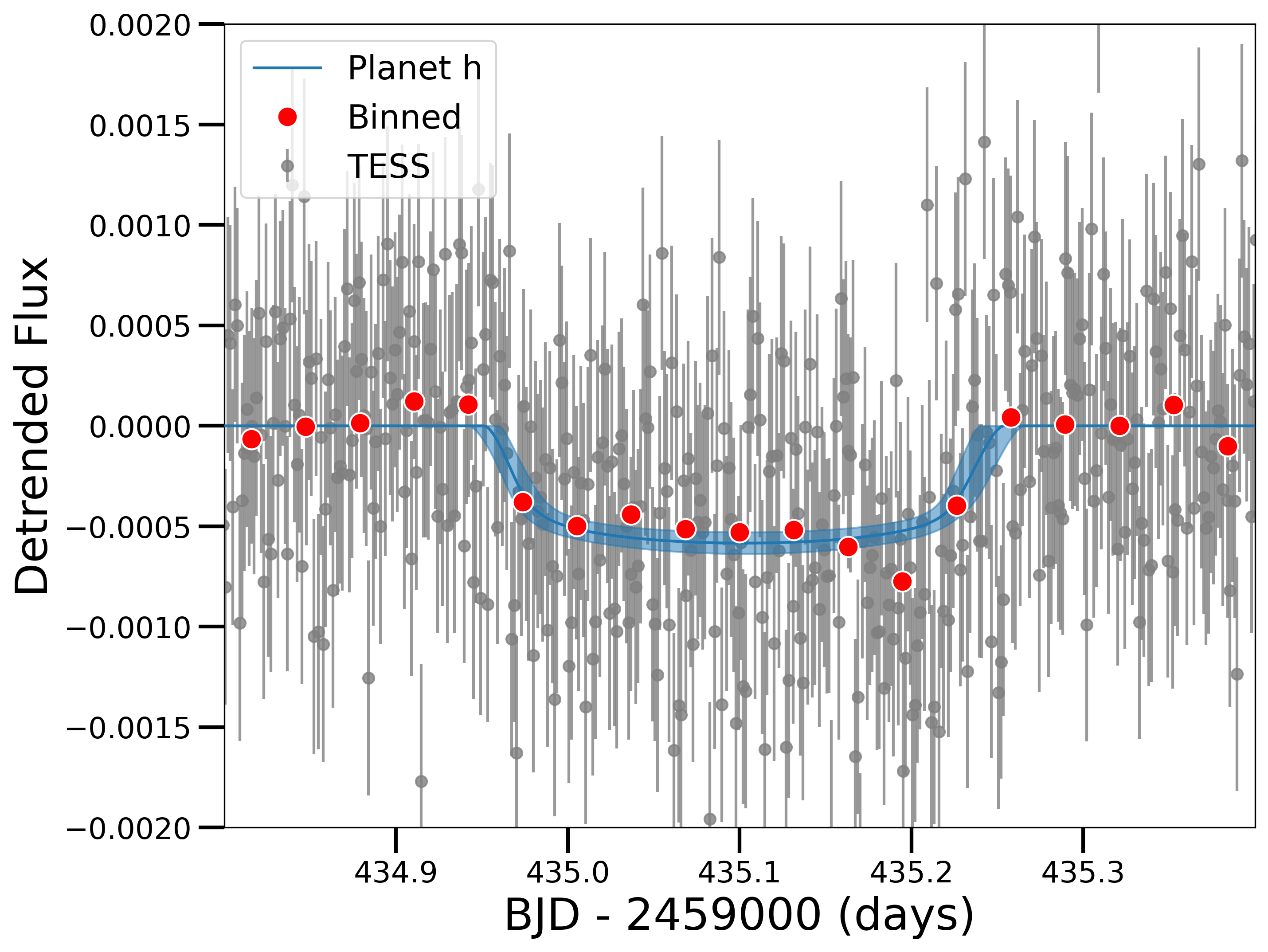}
    \caption{Our posterior transit fit to the single transit of the  candidate planet. Fits indicate that the planet likely has a radius near 2.68 R$_{\oplus}$. We use the SES (equation \ref{eqn:ses}) to verify the significance of the transit. }
    \label{fig:h_transit}
\end{figure}

We used \texttt{PyMC3} \citep{exoplanet:pymc3} to create a model context for the single transit of the candidate, and we used \texttt{starry} to generate the light curve model \citep{exoplanet:luger18}. \texttt{starry} uses a quadratic limb-darkening law when modeling transits. Eccentricity was modeled using a reparametrization detailed in \cite{exoplanet:vaneylen19}, and the orbital period and transit time were given normal priors from the posteriors of our nested sampling fits. Earlier fits were plagued by bimodal solutions related to transit depth, duration, and transit time. An in-transit region of slightly lower flux (visible in Figure \ref{fig:h_transit}) would sometimes confuse the model, shifting the transit time to the right and increasing the planet radius. To prevent this, we put a minimum transit duration of 0.2 days.

We then utilized a Hamiltonian Monte Carlo algorithm with the No-U-Turn Sampler \citep[NUTS;][]{hoffman11} to efficiently sample the posterior parameter space. We ran 4 chains, each with 5000 tuning steps and an additional 10000 parameter estimation steps. Our final fit is visible in Figure \ref{fig:h_transit}, and our posterior values are listed in Table \ref{tab:transit_post}.

\begin{deluxetable}{llll}
\label{tab:transit_post}
\tablecaption{TOI-1136(h) Transit Posteriors}
\tablehead{\colhead{~~~Parameter} &
\colhead{Posterior Value} & \colhead{Units} &\colhead{Description}
}
\startdata
 P$_{(h)}$ & 507$^{+303}_{-324}$ & Days & Orbital Period \\
 T$_{c}$ & 2459435.10$^{+0.006}_{-0.007}$ & BJD & Transit Time \\
R$_{p}$ & 2.68$^{+0.20}_{-0.18}$ & R$_{\oplus}$ & Planet Radius\\
e & 0.04$^{+0.05}_{-0.03}$ & ... & Eccentricity \\
 $\omega$ & 0$\pm$120 & Degrees & Arg. of Peri. \\
i$_{(h)}$ & 89.68$\pm$0.02 & Degrees & Inclination \\
t$_{14}$ & 0.26$^{+0.02}_{-0.01}$ & Days & Transit Duration \\
\enddata
\end{deluxetable}

The single-event nature of this possible planetary transit makes assessment of its veracity difficult. We use the single event statistic (SES, \citealt{jenkins02_a}) to quantify the quality of this candidate transit:

\begin{equation}
    \label{eqn:ses}
    \centering
    \mathrm{SES} = \frac{d \cdot s}{\sigma\sqrt{s^{T} s}}
\end{equation}

Above, $d$ is the detrended flux data, and $s$ is the predicted transit signal at each flux timestamp. $\sigma$ is the out-of-transit scatter. T indicates transposing a matrix. We use a subset of the detrended flux in D23, which removed correlated noise using a cubic spline of 0.5 days. Some correlated noise was still present even after such a detrending, especially near the wings of the transit. We mask the transit and additionally use a univariate spline with a smoothing factor of s=1400 to remove ther remainder of the out-of-transit variability. The SES is often expanded upon as a multi-event statistic (MES) in \textit{Kepler} systems \citep{twicken16}, though such an expansion is not possible in the case of a single transit. \cite{jenkins02_b} suggest 4.0 as a more conservative cutoff to call a single event significant, and 3.5 as sufficient. We estimate an SES of 12.3 for the single transit of this candidate planet, suggesting that the single transit is indeed statistically significant, and not likely due to white noise. We attempt to recover the mass of this candidate in the next section. 

Our conclusion then is that a single transit was most likely detected in the photometry that is not consistent with any of the known planets in the system, but any other parameters for this candidate planet are difficult to discern without a more in-depth analysis, or additional photometry. Additionally, we can rule out most false positive scenarios, as mentioned in D23.

\section{Analysis}
\label{sec:analysis}

The high multiplicity of the system generates a large number of free parameters in the model to describe the planetary orbits. The youth of the system suggests that large amounts of magnetic activity are likely occurring on the surface and within the star. This magnetic activity is likely to generate variability in the RVs and photometry not related to planetary motion. Indeed, examination of the quasi-periodic modulation of the TESS photometry confirms this expectation, and the high scatter of the RVs (RMS = 43.5 m s$^{-1}$) could not come from any of the known planets, even in the implausible event that they were all pure iron. Thus, some model to account for stellar variability in the RVs that is many times larger than the exoplanet signals is an essential part of modeling the RVs. 

A detailed photometric analysis of TOI-1136 was carried out in D23, including the identification of individual transit times at each transit epoch for each planet.  In this work, we jointly model the transit times determined in D23 and our newly collected RVs. 

We performed RV-only analyses of TOI-1136, but we failed to significantly detect the system's exoplanets for two reasons. First, the proximity of the stellar rotation period (8.4 days) to several planetary periods (P$_{b}$ = 4.17, P$_{c}$ = 6.26, P$_{d}$ = 12.52 days) made distinction challenging. Additionally, the significantly larger amplitude of the spot-induced variability ($\sim$ 50 m s$^{-1}$) compared to the expected planetary semi-amplitudes \citep[estimated from a mass-radius relationship; K$_{\rm{M-R}}$ = 0.3 - 3.0 m s$^{-1}$;][]{chen17} further hindered detection. TTV fits alone were much more successful at measuring planet masses, as the photometry is significantly easier to disentangle from stellar variability. It is expected that combining both TTV-predicted masses and RV-predicted masses would yield the best results, as the independent datasets can be combined in likelihood space to give the largest quantity of information about the system. We detail activity model training in \S \ref{sec:ffprime}, our complete RV + TTV model in \S \ref{sec:ttv_rv}, and detail our cross-validation in \S \ref{sec:cross validation}.

\subsection{Training the Activity Model}
\label{sec:ffprime}

\begin{figure}
    \centering
    \includegraphics[width=0.5\textwidth]{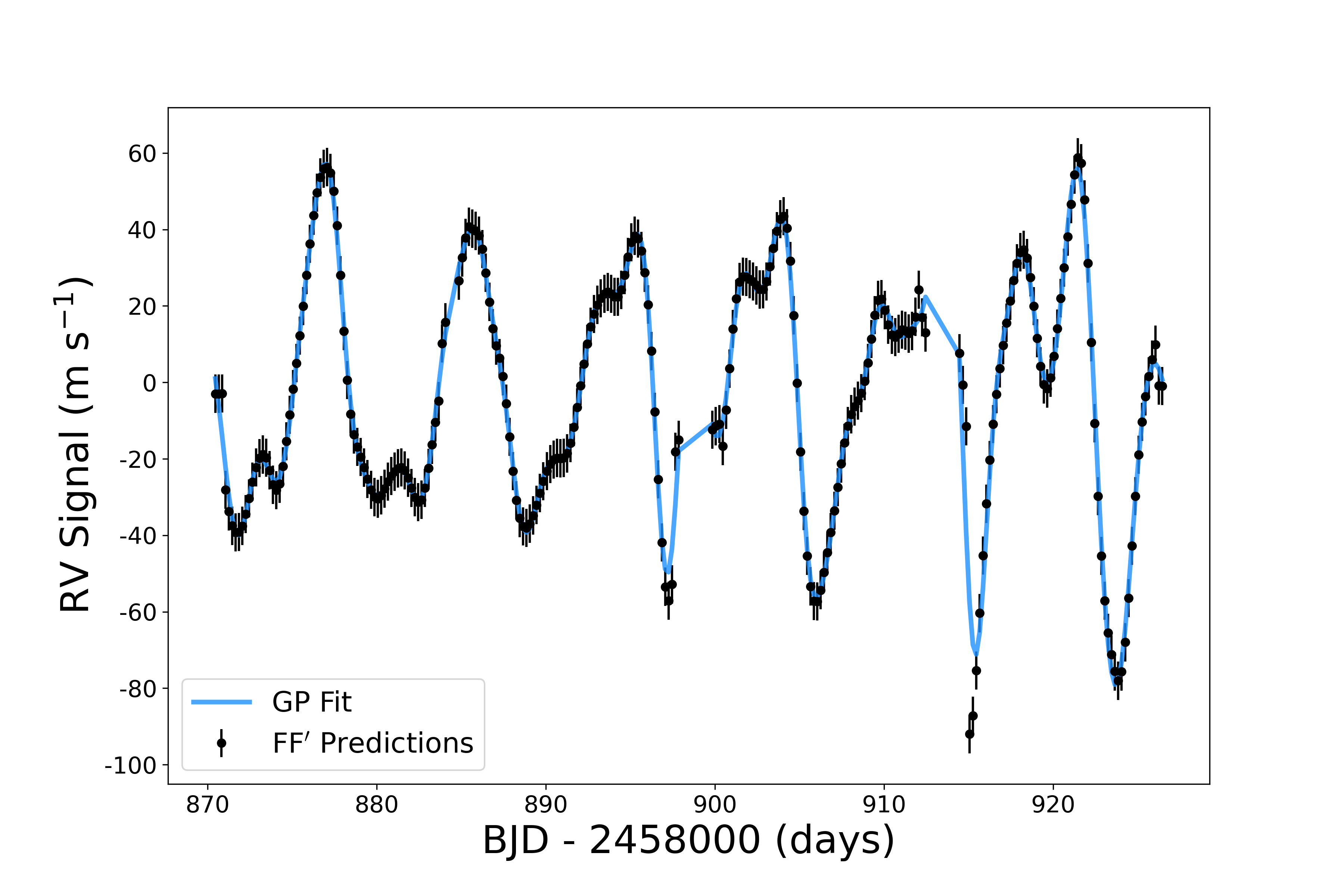}
    \caption{A plot of the GP fit to the RV activity signal of TOI-1136, calculated from photometry. This activity prediction is estimated via the FF$^{\prime}$ method described in \cite{aigrain12}. The above plot illustrates our fit to season 2. After assessing convergence, we use the frequency posteriors of this GP fit as priors on the GP hyperparameters for our TTV + RV fits in the next section.}
    \label{fig:s2_ffprime}
\end{figure}

We do not choose to include photometry in our final TTV + RV fit, but we can still use the 6 sectors of TESS data to inform our activity model. The RV contribution of the stellar activity can be predicted from photometry using the FF$^{\prime}$ method outlined in \cite{aigrain12}. This method is best at predicting quasi-periodic modulations from starspots or plage, which is likely the biggest contribution to TOI-1136's stellar activity. We fit the Quasi-Periodic kernel as described in \textsf{RadVel} documentation to the predicted RV activity signal \citep{fulton18}. We divide this signal into four ``seasons," corresponding to continuous TESS coverage. Season 1 is Sectors 14 and 15; season 2 is Sectors 21 and 22; season 3 is Sector 41; and season 4 is Sector 48. We then performed GP fits to each season individually, as well as a single run on all the FF$^{\prime}$ predictions together. An example plot of our fit to season 2 is visible in Figure \ref{fig:s2_ffprime}.

We performed an MCMC fit on the FF$^{\prime}$ data using \textsf{RadVel}, which assesses convergence by determining when the Gelman-Rubin (G-R) statistic is less than 1.01 and the number of independent samples is greater than 1000 for all free parameters for at least five consecutive checks \citep{ford06}.

The FF$^{\prime}$ spot model is relatively simple, and does not take into account all physical processes that occur in a magnetically active star. Additionally, it is known to break down in multi-spot cases. The model is based on photometry, which is expected to have a shared frequency structure with the RVs, though the phase may not be consistent. Consequently, we utilize only the posteriors of the terms associated with frequency ($\eta_{2}$,$\eta_{3},\eta_{4}$) as priors in our full TTV + RV model, and we maintain broad, uniform priors on GP amplitude terms.

\subsection{TTV + RV Model}
\label{sec:ttv_rv}

We used \textsf{TTVFast} \citep{deck14} to jointly model the transit times and the RVs of TOI-1136. \textsf{TTVFast} is a symplectic N-body integrator that uses Keplerian interpolation between N-body time steps to predict transit times \citep{deck14}. \textsf{TTVFast} uses seven free parameters per planet to integrate the dynamical motion of the system: planet mass, orbital period, eccentricity, argument of periastron, orbital inclination, mean anomaly at reference epoch, and the longitude of the ascending node. During our analysis of TOI-1136, we fixed the longitude of ascending node to zero for all planets, as our current data is not generally good at constraining this parameter, and this is commonly done \citep[e.g. D23,][]{grimm18}. All other parameters were left free to vary during our fits, resulting in six free parameters per planet. We perform fits on six and seven-planet models in order to better quantify the plausibility of including the candidate planet, as well as to examine the sensitivity of posteriors to including a seventh planet. Thus, there are 36 and 42 free parameters corresponding to Keplerian motion in each model.

We integrated using a timestep suggested by \cite{deck14} by dividing the shortest orbital period by 25. Consequently, our integration time step used was 0.125 days. After integration, we use the predicted transit times from the \textsf{TTVFast} model and the observed transit times (taken from D23) to calculate a likelihood associated with the TTVs ($\mathcal{L}_{TTV}$):

\begin{equation}
    \footnotesize
    \label{eqn:Lttv}
    \centering
    \mathrm{log}\mathcal{L}_{TTV} = -\frac{1}{2}\sum\limits_{j=1}^{n} \sum\limits_{i}\bigg(\frac{T_{ob,i,j} - T_{pr,i,j}}{\sigma_{ob,i,j}} \bigg)^{2} + \log(2\pi\sigma_{ob,i})
\end{equation}
where $i$ is the $i^\mathrm{th}$ transit of planet $j$, and $n$ is the number of transiting planets in the model.

The N-body integration performed by \textsf{TTVFast} models planet positions at each integration time step, which is sufficient to calculate the predicted radial velocity signal of the modeled system. The presence of additional, non-Keplerian signals in the RVs, most likely coming from spots or plage, requires the inclusion of an activity model. We utilize a GP to model the correlated noise of the stellar activity. A GP creates a covariance matrix from its kernel that models the covariance between each RV data point with each other data point. This is ideal for modeling the expected quasi-periodic behavior of the activity signal. This matrix can be used with the residuals of the planet fit to completely model the system. This is represented in the RV likelihood function in Equation \ref{Lrv}.

\begin{equation}
    \label{Lrv}
    \centering
    \mathrm{log}\mathcal{L}_{RV} = -\frac{1}{2}\big(r^{T}\mathcal{K}r + \log|\mathcal{K}| + N\log(2\pi)\big)
\end{equation}

Above, $\mathcal{K}$ is the covariance matrix of our GP, $N$ is the number of RV data points, and $r$ is a vector of residuals to the \textsf{TTVFast} predicted RV model given by Equation \ref{eqn:rvresid}.

\begin{equation}
    \label{eqn:rvresid}
    \centering
    r = RV_{obs} - RV_{pred} - \gamma
\end{equation}

Above, $\gamma$ corresponds to a linear offset subtracted from the model. A different offset is fit for each instrument, and subtracted from velocities of each instrument uniformly.

Our choice of GP kernel is the chromatic $\mathcal{K}_{J1}$ kernel outlined in \cite{cale21}. This GP kernel is an expansion of the commonly used Quasi-Periodic GP kernel \citep{haywood14,lopezmorales16}. The $\mathcal{K}_{J1}$ kernel utilizes a different amplitude parameter for each instrument used in the fit, which is particularly useful for RV instruments of different wavelength regimes, as stellar activity is expected to be chromatic \citep{crockett12,robertson20}. This is not highly relevant in the case of TOI-1136, as the central wavelength bands of all three instruments used (HIRES, APF, HARPS-N) are close in wavelength-space (Though HARPS-N is not an iodine instrument, and this might have a significant effect). However, this $\mathcal{K}_{J1}$ kernel can be used to model all three instruments simultaneously in one covariance matrix, rather than the traditional method of calculating a likelihood for each instrument and summing them. Consequently, RVs from each instrument maintain a covariance even between RVs of other instruments. This is particularly useful for preventing overfitting of the GP, which is a serious problem, especially in a model with so many free parameters. This is discussed more in \S \ref{sec:cross validation}. 

Our total joint model log likelihood is

\begin{equation}
    \label{eqn:Ltotal}
    \centering
    \mathrm{log}\mathcal{L}_{tot} = \mathrm{log}\mathcal{L}_{TTV} + \mathrm{log}\mathcal{L}_{RV} + \mathrm{log}\mathcal{P},
\end{equation}

where $\mathcal{P}$ is the product of all priors.

We generally adopted broad priors on the free parameters of TOI-1136, with a few exceptions. The inclinations, while constrained by TTVs, are also informed by transit shapes, which we did not fit for in our model, but which were fit in D23. To leverage this information without including transit fits in our model, we use a Gaussian prior on the inclination of the inner 6 planets, with values corresponding to the posteriors of D23. To prevent the perfect degeneracy between inclinations on either side of 90 degrees, we also put an upper limit of 90$^\circ$ on all the inclination priors, preventing chains from crossing that threshold. Technically, we are restricting mutual inclinations between planets to minimum values, when two planets could have inclinations on either side of the 90 degree threshold but still exhibit the same transit shape. However, this difference in mutual inclination is limited to only a few degrees, and is unlikely to affect our fit results, so we ignore it. We estimate the minimum and maximum inclinations possible for the candidate planet to transit, and use these values as uniform priors for TOI-1136 (h). The GP hyperparameters, too, can be informed by the photometry. This is particularly important due to the flexibility of GPs, and our model's susceptibility to overfitting. Uninformative priors on GP terms give the GP the flexibility to modify the model until residual scatter is minimized, even if the results are unphysical. We use the posteriors of a GP fit to the FF$^\prime$ predictions, detailed in \S \ref{sec:ffprime}, as priors on the GP hyperparameters. A full list of our priors is visible in table \ref{tab:priors}.

\begin{deluxetable*}{lllll}
\label{tab:priors}
\tabletypesize{\footnotesize}
\tablecaption{Priors Used for Various Fits}
\tablehead{\colhead{~~~Parameter Name} &
\colhead{TTVFast Prior} & \colhead{FF$^{\prime}$ Prior} &  \colhead{Units} & \colhead{Description}
}
\startdata
\sidehead{\textbf{Planet Priors (b-g):}}
~~~P$_{orb}$ & $\mathcal{U}^{a}(0.99*\mu_{D23}^{*}, 1.01*\mu_{D23})$ & - & days & Period \\
~~~$\sqrt{e} \cos\omega$ & $\mathcal{U}(-1, 1)$  & - &  ...  & Eccentricity Reparametrization \\
~~~$\sqrt{e} \sin\omega$ & $\mathcal{U}(-1, 1)$ & - &  ...  & Eccentricity Reparametrization \\
~~~$\frac{m_{p}}{m_{s}}$ & $\mathcal{U}(0.0,0.01)$ & - & ... & Planet-star Mass Ratio \\
~~~$M$ & $\mathcal{U}(-180,180)$ & - &  degrees & Mean Anomaly \\
~~~$i$ & $\mathcal{BN}^{b}(\mu_{D23}, sd_{D23}, 80, 90)$ & - &  degrees  & Inclination \\
\sidehead{\textbf{TOI-1136(h) Priors:}}
~~~P$_{orb}$ & $\mathcal{U}(1, 1000)$ & - &  days & Period \\
~~~$\sqrt{e} \cos\omega$ & $\mathcal{U}(-1, 1)$  & - & ...  & Eccentricity Reparametrization \\
~~~$\sqrt{e} \sin\omega$ & $\mathcal{U}(-1, 1)$ & -  & ...  & Eccentricity Reparametrization \\
~~~$\frac{m_{p}}{m_{s}}$ & $\mathcal{U}(0.0,0.01)$ & - & ... & Planet-star Mass Ratio \\
~~~$M$ & $\mathcal{U}(-180,180)$ & - &  degrees & Mean Anomaly \\
~~~$i$ & $\mathcal{U}(89.5, 90)$ & - &  degrees  & Inclination \\
\sidehead{\textbf{GP Hyperparameters}}
~~~$\eta_{1, \rm{HIRES}}$ & $\mathcal{U}(1,100)$ & - &  m s$^{-1}$  & HIRES GP Amplitude \\
~~~$\eta_{1, APF}$ & $\mathcal{U}(1,100)$ & - & m s$^{-1}$ & APF GP Amplitude \\
~~~$\eta_{1, HARPS-N}$ & $\mathcal{U}(1,100)$ & - & m s$^{-1}$   & HARPS-N GP Amplitude \\
~~~$\eta_{1, FF^{\prime}}$ & - & $\mathcal{J}^{c}(0.001,100.0)$ &  m s$^{-1}$   & FF$^{\prime}$ GP Amplitude \\
~~~$\eta_{2}$ & $\mathcal{N}^{d}(9.6188,0.871)$ & $\mathcal{J}(8.54,10^{10})$ & days   & Exponential Scale Length \\
~~~$\eta_{3}$ & $\mathcal{N}(8.429,0.094)$  & $\mathcal{N}(8.429,0.094)$  &  days   & Periodic Term \\
~~~$\eta_{4}$ & $\mathcal{N}(0.4402,0.0499)$  & $\mathcal{J}(10^{-5},1)$ &  ...  & Periodic Scale Length \\
\sidehead{\textbf{Instrumental Parameters}}
~~~$\gamma_{\rm{HIRES}}$ & $\mathcal{U}(-100,100)$ & - &  m s$^{-1}$  & HIRES offset \\
~~~$\gamma_{\rm{APF}}$ & $\mathcal{U}(-100,100)$ & - &  m s$^{-1}$ & APF offset \\
~~~$\gamma_{\rm{HARPS-N}}$ & $\mathcal{U}(-100,100)$ & - &  m s$^{-1}$   & HARPS-N offset \\
~~~$\gamma_{\rm{FF^{\prime}}}$ & ... & $\mathcal{U}(-100,100)$ &  m s$^{-1}$   & FF$^{\prime}$ offset \\
~~~$\sigma_{\rm{HIRES}}$ & $\mathcal{U}(0.01,100)$ & - &  m s$^{-1}$    & Instrumental Jitter, HIRES \\
~~~$\sigma_{\rm{APF}}$ & $\mathcal{U}(0.01,100)$ & - &  m s$^{-1}$   & Instrumental Jitter, APF \\
~~~$\sigma_{\rm{HARPS-N}}$ & $\mathcal{U}(0.01,100)$ & - &  m s$^{-1}$    & Instrumental Jitter, HARPS-N \\
~~~$\sigma_{\rm{FF^{\prime}}}$ & - & $\mathcal{U}(0.01,100)$ &  m s$^{-1}$ & Instrumental Jitter, FF$^{\prime}$ \\
\enddata
\tablenotetext{a}{$\mathcal{U}$ is a uniform prior with $\mathcal{U}$(lower,upper)}
\tablenotetext{b}{$\mathcal{BN}$ is a bounded normal prior with $\mathcal{BN}$(mean, standard deviation, minimum, maximum)}
\tablenotetext{c}{$\mathcal{J}$ is a Jeffreys prior with $\mathcal{J}$(lower,upper)}
\tablenotetext{d}{$\mathcal{N}$ is a normal prior with $\mathcal{N}$(mean, standard deviation)}
\tablenotetext{*}{$\mu_{D23}$ refers to the mean posterior taken from Table 10 in D23. $sd_{D23}$ refers to the 1$\sigma$ uncertainty taken from Table  10 in D23.}
\tablenotetext{-}{ indicates a free parameter that was not fit in that particular model.}
\end{deluxetable*}

We initially perform a simple least-squares optimization on our model using \texttt{lmfit} \citep{newville14}. We let all parameters vary during this optimization step, except for the GP hyperparameters which are fixed. This is partially to prevent some measure of overfitting, which a least-squares optimization may do for a complicated model, and is additionally unceccessary: our FF$^{\prime}$ fits detailed in \S \ref{sec:ffprime} already provide a good estimate of our GP hyperparameter values, and their uncertainties.

To model the posterior probability of our TTV + RV model, we used the \texttt{emcee} software package to perform Markov-Chain Monte Carlo (MCMC) inference \citep{foremanmackey13}. We utilize Differential Evolution MCMC (DEMCMC) sampling with the DEMove in \texttt{emcee} documentation \citep{terbraak06} as well as the affine-invariant sampler proposed in \cite{goodman10} for faster MCMC convergence, referred to in \texttt{emcee} documentation as the StretchMove. We experimented with different hyperparameter values to tune the sampling, and we settled on sigma=2e-8 and gamma0=0.33 for the DEMove, as this combination produced the desired acceptance rate near 30$\%$. We set the single hyperparameter for the StretchMove to a=1.2, as this value produced the highest acceptance. Both methods produced consistent results, though we report our results from the DEMove.

We estimated convergence via the method proposed in \cite{goodman10} and further endorsed in \cite{foremanmackey13}, by estimating the integrated autocorrelation time, $\tau$. This value is approximated by \texttt{emcee} during the MCMC process, and the estimate asymptotically approaches the correct value as more steps in the sampling are computed. \texttt{emcee} documentation suggests using a large number of simultaneous walkers, or chains, to more efficiently sample parameter-space, and to more accurately estimate $\tau$. The sampler should be run for multiple lengths of $\tau$ to ensure that final results are not subject to sampler uncertainty, and that final results sufficiently reflect measured uncertainties of the data and model. 

Less complicated models that utilized \texttt{TTVFast} in the literature were able to achieve precise results using only dozens of walkers and tens of thousands of sampler steps \citep[e.g.][]{becker15,tran23}. Due to its increased complexity, however, we utilized 1000 simultaneous walkers for TOI-1136, and we ran for 300,000 MCMC steps for both models. Our models estimate $\tau$ at $\sim 25000$ model steps, suggesting that our model has run for $>$ 10 autocorrelation timescales. We also compute the G-R statistic to compare inter-chain and intra-chain variability \citep{ford06}. Our model meets convergence criteria (G-R $<$ 1.01 for all parameters) according to the G-R statistic, though we caution that the this is considered less robust than autocorrelation times when using the StretchMove ensemble in \textsf{emcee}.

Our final posterior parameter values are listed in Table \ref{tab:posteriors}. A plot of our RVs and TTVs modeled to these values is visible in Figures \ref{fig:RV} and \ref{fig:TTV}. To encourage reproduction, we provide a public github repository with our analysis code and encourage others to use and test it\footnote{\url{https://github.com/CCBeard/TOI-1136_Analysis_Code} \citep{beard23_zenodo}}.

Beyond the TTV + RV models described above, we ran a TTV-only model as well. This will help us to quantify the effect RVs are having on our models more directly, and additionally help when comparing results with D23. We only performed such a fit for a six planet model, as a TTV-only fit with a single transiting planet is not highly meaningful. These results are reported in Table \ref{tab:6pttvonly}, and are discussed further in \S \ref{sec:discussion}.

\subsection{Cross Validation}
\label{sec:cross validation}

Our joint model (described in \S \ref{sec:ttv_rv}) has a large number of free parameters with respect to the size of the dataset, and is consequently susceptible to overfitting. In principle, a dataset is overfit when it learns the training data too well, and starts to recreate the statistical noise of the data in its predictions, rather than information about a physical system. When training a physically motivated model on data, the training likelihood of the model should initially improve as the model learns the features of the data. However, the training likelihood will often continue to improve (as the model learns the noise properties of the data it sees), as its predictive ability on data it doesn't see (the test dataset) begins to fall. When the model likelihood increases at the expense of predictivity, we call this overfitting. We are most interested in determining whether our final hyperparameters from the model in \S \ref{sec:ttv_rv} are contributing to overfitting, and our intention is not to estimate model parameters in this section.

We perform cross validation to assess our model's predictive ability on data it has not seen before. Ideally, we would follow the method proposed in \cite{blunt23}, reserving 30$\%$ of our radial velocity data as a ``test dataset" and only training our model on 70$\%$. We could, at fixed intervals, check our model's predictive ability and determine when the test likelihood starts worsening. This method is not ideal for TOI-1136, however, for a number of reasons. Despite our large model with 52 free parameters, our actual dataset contains a relatively small number of points (87 transit times, 410 RVs, 497 datapoints). Removing 30$\%$ of our dataset would largely reduce the size of a dataset already worryingly close to the number of free parameters. Furthermore, shrinking this percentage would likely result in a test dataset that is not well representative of the whole. Lastly, such tests often work best when repeated many times to ensure that the randomly drawn test dataset is representative of the whole sample. Our models are already extremely expensive to run (taking around 5000 CPU-hours to converge), and repeating them dozens or hundreds of times would be prohibitively so. Additionally, our model requires large amounts of random access memory (RAM) to manipulate the long (300,000 steps) and wide (1000 chains) samples object, and our access to specialized high-memory CPUs is additionally limited.  Because of these constraints, we make a compromise between a simpler cross-validation utilized in \cite{hara20} and the more complicated method utilized in \cite{blunt23}.

\cite{hara20} utilize cross-validation of their GP model by creating a grid of GP hyperparameter values, and optimizing planet models with GP hyperparameters fixed at these values. This optimization is performed on 70$\%$ of the data, and the authors then evaluate the likelihood of the 30$\%$ test dataset.

When applying this to TOI-1136, we focus on the hyperparameters of the $\mathcal{K}_{J1}$ GP kernel. GP parameters often cause overfitting, as GPs are incredibly flexible. Because the rotation period of TOI-1136 is clearly detected in \S \ref{sec:acf}, we focus only on the parameters $\eta_{1}$, $\eta_{2}$, and $\eta_{4}$, known as the GP amplitude, exponential decay length, and periodic scale length, respectively (described in \cite{dai17}). We perform a grid search using these three parameters, performing fits with them fixed at certain values. We prioritize values distributed around the posterior of our model. The amplitude term was tested at values of 10, 50, 75, and 100 m s$^{-1}$. The exponential decay length was allowed values of 5, 10, 50, 100, 1000. Finally, the periodic scale length was tested with values of 0.1, 0.2, 0.3, 0.4, 0.5. For comparison, our final 7 planet model values, listed in Table \ref{tab:posteriors}, are $\eta_{1,HIRES}$ = 36.9 m s$^{-1}$, $\eta_{1,APF}$ = 43.8 m s$^{-1}$, $\eta_{1,HARPS-N}$ = 37.4 m s$^{-1}$, $\eta_{2}$ = 13.5 days, and $\eta_{4}$ = 0.25.

\begin{figure}
    \centering
    \includegraphics[width=0.5\textwidth]{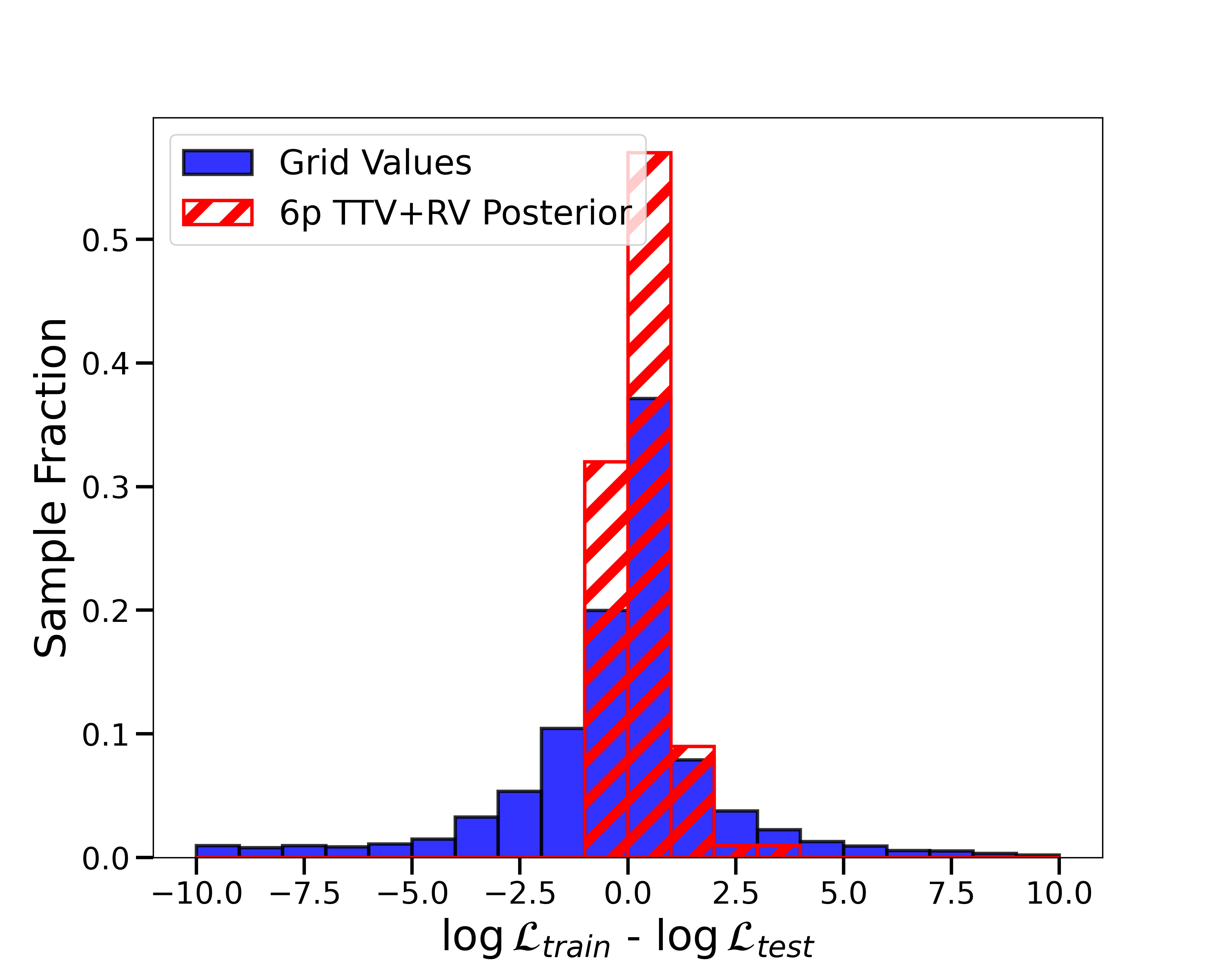}
    \caption{Histogram of the difference between the scaled log-likelihood of the training datasets described in \S \ref{sec:cross validation} and the scaled test datasets. As a model becomes overfit to data, the training likelihood should become much higher than the test likelihood. Positive values indicate overfitting, and negative values indicate underfitting. A histogram of our grid search results is shown in blue, and a histogram of the 100 samples using our 6 planet TTV+RV posterior are shown in red hatches. Our model likelihood differences skew slightly into overfitting, but they are much more highly concentrated around 0 than our grid search.}
    \label{fig:CV}
\end{figure}

\begin{figure*}
    \centering
    \includegraphics[width=\textwidth]{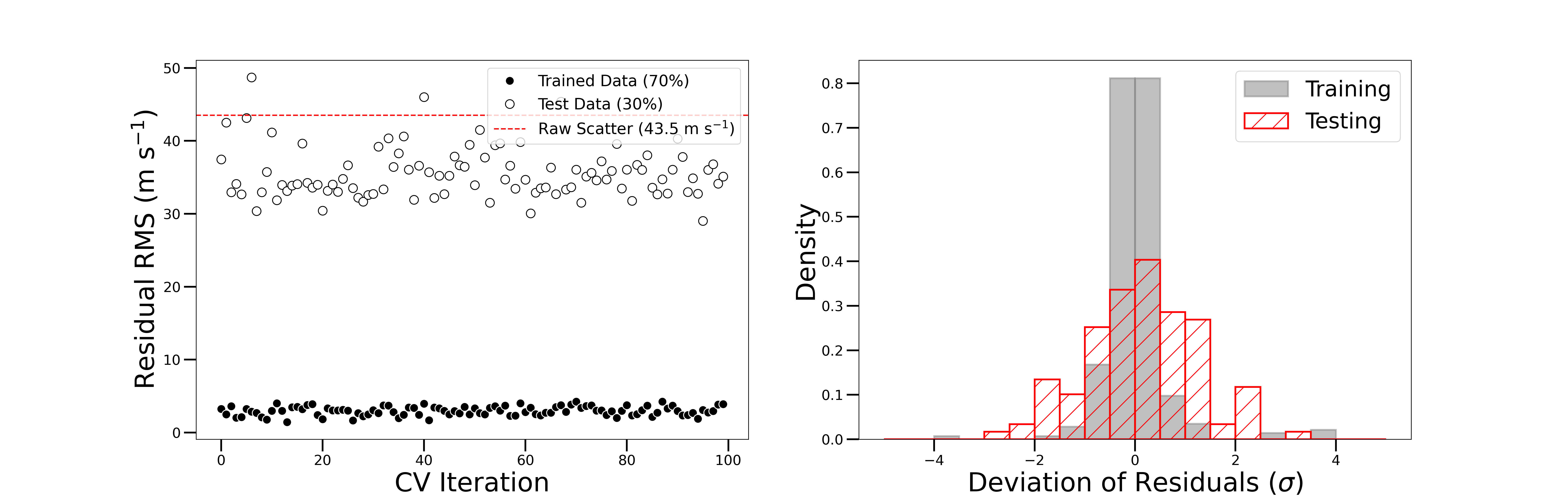}
    \caption{Left: A comparison of the RMS of the training and test residuals during our cross validation. The test datasets exhibit higher scatter, but their GP uncertainty is also high, indicating that the model has trouble predicting the held-out, test dataset. Right: a histogram of the residuals/uncertainty of the training and test datasets. Despite the large RMS of the test dataset, its associated uncertainty is also high. Consequently, the predictions are not unreliable, though imprecise. This again indicates slight overfitting, though it is less than in \cite{blunt23}.}
    \label{fig:rms_cv}
\end{figure*}

For each combination of hyperparameter values, we set up the model described in \S \ref{sec:ttv_rv}, except with these three hyperparameters fixed at their selected value. We then split the RV data into a training dataset (70$\%$) and test dataset (30$\%$) randomly. We do so by instrument so that there is always the same number of APF, HIRES, and HARPS-N points. We then optimize the free parameters of the model using \textsf{lmfit} \citep{newville14} on the training dataset only. After optimization, we estimate the likelihood of the training and test datasets. The training and test datasets are then recreated again via random draw, and this process is repeated 100 times for each parameter trio, and we take the average result.  Note that we only perform a least-squares optimization to the model, rather than a full MCMC. This is to prevent this check from being prohibitively expensive.

We perform the same calculation with the three hyperparameters fixed at the values taken from our full MCMC posteriors. With a representative test likelihood value and training likelihood value for each combination of GP hyperparameters and for our final model, we scale each likelihood by the number of datapoints used to estimate. The result is a variety of likelihoods that can be compared on the same scale. The idea is that if the training likelihood is much better than the test likelihood, the model is probably overfitting. Ideally, the scaled test and training likelihoods should be close to the same value, suggesting that the model makes predictions on both datasets equally well.

We additionally follow \cite{blunt23} and check the predictivity of our activity model. We do so by measuring scatter of the residuals of our training datasets compared to the scatter of the residuals of the test datasets.

Our final results in Figure \ref{fig:CV} suggest that our final values might be slightly overfitting the data, but that the posterior GP hyperparameter values are well within a normal range, and most of our models are concentrated around zero (neither under or over-fitting). In Figure \ref{fig:rms_cv}, we see the difference in residual RMS between the training and test dataset. The test dataset exhibits much higher scatter than the training dataset, which is expected, but may indicate overfitting considering the degree of difference. We recreate Figure 2 from \cite{blunt23} by picking an iteration at random and plotting a histogram of the prediction value divided by its uncertainty (1$\sigma$ GP standard deviation + RV error, added in quadrature). While the test dataset exhibits higher scatter than the training dataset, it is not as anomalous as the result demonstrated in \cite{blunt23}. This result agrees with the earlier likelihood estimation that our model may be slightly overfit, but that it is not likely to be extreme.

Overfitting might alter posterior parameters into unphysical regimes, or cause under or over-estimated posterior errors. Our general median agreement with D23 suggest that the former is not likely happening, though our improved errors over D23 suggest that error under-estimation may be happening. A combined RV/TTV analysis is expected to improve precision, however, and MCMC convergence checks, as well as our cross validation above suggest that if this is happening, it is not extreme.

%\FloatBarrier

\startlongtable
\begin{deluxetable*}{lllll}
\centering
\tablecaption{TTV+RV Posteriors of TOI-1136$^{\dag}$}
\label{tab:posteriors}
\tablehead{\colhead{~~~Parameter Name} & 
\colhead{6p TTV+RV Posterior} & \colhead{7p TTV+RV Posterior} & \colhead{Units} & \colhead{Description}}
\startdata
\sidehead{\textbf{Planet b}}
\hline
\sidehead{~~\textit{Fit Parameters}}
~~~P$_{b}$ &  4.1727 $\pm$ 0.0003 & 4.1728$^{+0.0003}_{-0.0002}$ & days & Orbital Period \\
~~~$\sqrt{e} \cos\omega_{b} $ & 0.15$^{+0.02}_{-0.03}$ & 0.18$^{+0.02}_{-0.03}$ &  ... & Eccentricity Reparametrization \\
~~~$\sqrt{e} \sin\omega_{b} $ &  0.07$^{+0.03}_{-0.04}$ & 0.04$\pm$0.06 & ... & Eccentricity Reparametrization \\
~~~ M$_{b}$ &  51.9$^{+12.0}_{-10.3}$ & 64.9$^{+17.7}_{-18.7}$ & degrees & Mean Anomaly \\
~~~i$_{b}$ & 86.4$\pm$0.6 & 86.4$\pm$0.3 &  degrees  & Inclination \\
~~~m$_{p,b}$ & 3.50$^{+0.8}_{-0.7}$  & 3.68$^{+0.61}_{-0.54}$ & M$_{\oplus}$ & Planet Mass \\
\sidehead{~~\textit{Derived Parameters}}
~~~ $\rho_{b}$ & 2.80$\pm$1.00  & 2.95$\pm$0.96 & g/cc & Bulk Density \\
~~~ e$_{b}$ & 0.027$\pm$0.009 & 0.03$\pm$0.01 & ... & Eccentricity \\
~~~ $\omega_{b}$ & 25$\pm$11 & 12.5$\pm$9 & Degrees & Argument of Periastron \\
~~~ K$_{b}^{*}$ &  1.37$\pm$0.29 & 1.44$\pm$0.22 & m s$^{-1}$ & RV Semi-amplitude \\
~~~ a$_{b}$ &  0.05106$\pm$0.0009 & 0.0511$\pm$0.0008 & AU & Semi-Major Axis \\
~~~ T$_{eq,b}^{**}$ &  1216$\pm$12 & 1216$\pm$11 & K & Equilibrium Temperature \\
\sidehead{\textbf{Planet c}}
\hline
\sidehead{~~\textit{Fit Parameters}}
~~~P${_c}$ & 6.2574$\pm$0.0002 & 6.2577$^{+0.0003}_{-0.0002}$ & days & Orbital Period \\
~~~$\sqrt{e} \cos\omega_{c} $ &  -0.11$\pm$0.01 & -0.08$\pm$0.02 &  ... & Eccentricity Reparametrization \\
~~~$\sqrt{e} \sin\omega_{c} $ & -0.31$\pm$0.02 & -0.29$\pm$0.02 & ... & Eccentricity Reparametrization \\
~~~ M$_{c}$ & 62.9$^{+2.2}_{-2.4}$ & 57.3$^{+3.6}_{-3.5}$ &  degrees & Mean Anomaly \\
~~~i$_{c}$ & 88.8$^{+0.7}_{-1.0}$ & 89.3$^{+0.5}_{-0.4}$ &   degrees  & Inclination \\
~~~m$_{p,c}$ & 6.32$^{+1.1}_{-1.3}$ & 7.41$^{+0.98}_{-1.20}$ &  M$_{\oplus}$ & Planet Mass \\
\sidehead{~~\textit{Derived Parameters}}
~~~ $\rho_{c}$ & 1.45$\pm$0.29 & 1.71$\pm$0.28 & g/cc & Bulk Density \\
~~~ e$_{c}$ & 0.11$\pm$0.01 & 0.09$\pm$0.01 & ... & Eccentricity \\
~~~ $\omega_{c}$ & 70$\pm$2 & 74$\pm$4 & Degrees & Argument of Periastron \\
~~~ K$_{c}$ & 2.16$\pm$0.41 & 2.54$\pm$0.38 & m s$^{-1}$ & RV Semi-amplitude \\
~~~ a$_{c}$ & 0.0669$\pm$0.0005 & 0.0669$\pm$0.0007 & AU & Semi-Major Axis \\
~~~ T$_{eq,c}$ & 1062$\pm$7 & 1062$\pm$8 & K & Equilibrium Temperature \\
\sidehead{\textbf{Planet d}}
\hline
\sidehead{~~\textit{Fit Parameters}}
~~~P$_{d}$ & 12.5199$\pm$0.0004 & 12.5195$^{+0.0003}_{-0.0004}$ &  days & Orbital Period \\
~~~$\sqrt{e} \cos\omega_{d} $ & -0.10$\pm$0.01  & -0.07$\pm$0.02 &  ... & Eccentricity Reparametrization \\
~~~$\sqrt{e} \sin\omega_{d} $ & 0.10$\pm$0.01 & 0.18$\pm$0.02 &  ... & Eccentricity Reparametrization \\
~~~ M$_{d}$ & 140.6$^{+3.8}_{-3.4}$ & 165.0$^{+5.1}_{-5.0}$ &  degrees & Mean Anomaly \\
~~~i$_{d}$ & 89.2$\pm$0.5 & 89.4$\pm$0.3 &  degrees  & Inclination \\
~~~m$_{p,d}$ & 8.35$^{+1.8}_{-1.6}$ & 5.6$^{+0.9}_{-1.0}$ & M$_{\oplus}$ & Planet Mass \\
\sidehead{~~\textit{Derived Parameters}}
~~~ $\rho_{d}$ & 1.81$\pm$0.35 & 0.31$\pm$0.06 & g/cc & Bulk Density \\
~~~ e$_{d}$ & 0.042$\pm$0.004 & 0.04$\pm$0.01 & ... & Eccentricity \\
~~~ $\omega_{d}$ & -67$\pm$3 &  -68$\pm$6 & Degrees & Argument of Periastron \\
~~~ K$_{d}$ & 2.27$\pm$0.46 & 1.52$\pm$0.27 & m s$^{-1}$ & RV Semi-amplitude \\
~~~ a$_{d}$ & 0.1062$\pm$0.0008 & 0.1062$\pm$0.0007 & AU & Semi-Major Axis \\
~~~ T$_{eq,d}$ & 843$\pm$6 & 843$\pm$5 & K & Equilibrium Temperature \\
\sidehead{\textbf{Planet e}}
\hline
\sidehead{~~\textit{Fit Parameters}}
~~~P$_{e}$ & 18.801$\pm$0.001 & 18.802$\pm$0.001 & days & Orbital Period \\
~~~$\sqrt{e} \cos\omega_{e} $ & 0.08$\pm$0.01  & 0.08$\pm$0.01 & ... & Eccentricity Reparametrization \\
~~~$\sqrt{e} \sin\omega_{e} $ & -0.19$\pm$0.01 & -0.22$\pm$0.01 & ... & Eccentricity Reparametrization \\
~~~ M$_{e}$ & 175.5$^{+2.1}_{-2.2}$ & 175.5$^{+2.8}_{-3.5}$ & degrees & Mean Anomaly \\
~~~i$_{e}$ & 89.2$\pm$0.5 & 89.3$\pm$0.3 &  degrees  & Inclination \\
~~~m$_{p,e}$ & 6.07$^{+1.09}_{-1.01}$ & 3.31$^{+0.46}_{-0.39}$ & M$_{\oplus}$ & Planet Mass \\
\sidehead{~~\textit{Derived Parameters}}
~~~ $\rho_{e}$ & 1.81$\pm$0.35 & 0.99$\pm$0.16 & g/cc & Bulk Density \\
~~~ e$_{e}$ & 0.0425$\pm$0.004 & 0.0548$\pm$0.005 & ... & Eccentricity \\
~~~ $\omega_{e}$ & -67$\pm$3 & -70$\pm$2.4 & Degrees & Argument of Periastron \\
~~~ K$_{e}$ & 1.44$\pm$0.25 & 0.78$\pm$0.102 & m s$^{-1}$ & RV Semi-amplitude \\
~~~ a$_{e}$ &  & 0.139$\pm$0.002 & AU & Semi-Major Axis \\
~~~ T$_{eq,e}$ & 737$\pm$6 & 736$\pm$6 & K & Equilibrium Temperature \\
\sidehead{\textbf{Planet f}}
\hline
\sidehead{~~\textit{Fit Parameters}}
~~~P$_{f}$ & 26.321$\pm$0.001 & 26.3213$\pm$0.001 & days & Orbital Period \\
~~~$\sqrt{e} \cos\omega_{f} $ &  -0.02$\pm$0.01  & -0.04$\pm$0.02 & ... & Eccentricity Reparametrization \\
~~~$\sqrt{e} \sin\omega_{f} $ & 0.02$\pm$0.01 & 0.05$\pm$0.02 & ... & Eccentricity Reparametrization \\
~~~ M$_{f}$ & 51.4$^{+4.3}_{-4.4}$ & 52.4$^{+10.1}_{-9.2}$ & degrees & Mean Anomaly \\
~~~i$_{f}$ & 89.3$\pm$0.4 & 89.4$^{+0.2}_{-0.3}$ &  degrees  & Inclination \\
~~~m$_{p,f}$ & 9.7$^{+3.9}_{-3.7}$ & 8.22$^{+2.8}_{-2.4}$ & M$_{\oplus}$ & Planet Mass \\
\sidehead{~~\textit{Derived Parameters}}
~~~ $\rho_{f}$ & 0.89$\pm$0.22 & 0.77$\pm$0.25 & g/cc & Bulk Density \\
~~~ e$_{f}$ & 0.001$\pm$0.001 & 0.0$\pm$0.003 & ... & Eccentricity \\
~~~ $\omega_{f}$ & -45$\pm$20 & -51$\pm$18 & Degrees & Argument of Periastron \\
~~~ K$_{f}$ & 2.01$\pm$0.46 & 1.74$\pm$0.55 & m s$^{-1}$ & RV Semi-amplitude \\
~~~ a$_{f}$ & 0.174$\pm$0.002 & 0.174$\pm$0.002 & AU & Semi-Major Axis \\
~~~ T$_{eq,f}$ & 658$\pm$5 & 658$\pm$5 & K & Equilibrium Temperature \\
\sidehead{\textbf{Planet g}}
\hline
\sidehead{~~\textit{Fit Parameters}}
~~~P$_{g}$ & 39.545$\pm$0.002 & 39.544$^{+0.001}_{-0.002}$ & days & Orbital Period \\
~~~$\sqrt{e} \cos\omega_{g} $ & 0.03$\pm$0.01  & 0.02$\pm$0.01 & ... & Eccentricity Reparametrization \\
~~~$\sqrt{e} \sin\omega_{g} $ & -0.19$\pm$0.02 & -0.20$\pm$0.02 & ... & Eccentricity Reparametrization \\
~~~ M$_{g}$ & -119.5$^{+2.3}_{-2.5}$& -118.5$^{+3.0}_{-2.6}$ & degrees &  Mean Anomaly \\
~~~i$_{g}$ & 89.5$\pm$0.3 & 89.7$\pm$0.2 &  degrees  & Inclination \\
~~~m$_{p,g}$ & 5.6$^{+4.1}_{-3.2}$ & 12.0$^{+5.2}_{-3.2}$ & M$_{\oplus}$ & Planet Mass \\
\sidehead{~~\textit{Derived Parameters}}
~~~ $\rho_{g}$ & 1.9$\pm$1.3 & 4.07$\pm$1.52 & g/cc & Bulk Density \\
~~~ e$_{g}$ & 0.04$\pm$0.01 & 0.04$\pm$0.01 & ... & Eccentricity \\
~~~ $\omega_{g}$ & -81$\pm$3 & -84$\pm$3  & Degrees & Argument of Periastron \\
~~~ K$_{g}$ & 1.03$\pm$0.68 & 2.22$\pm$0.78 & m s$^{-1}$ & RV Semi-amplitude \\
~~~ a$_{g}$ & 0.229$\pm$0.003 & 0.229$\pm$0.002 & AU & Semi-Major Axis \\
~~~ T$_{eq,g}$ & 574$\pm$5 & 574$\pm$4 & K & Equilibrium Temperature \\
\sidehead{\textbf{TOI-1136 (h)$^{***}$}}
\hline
\sidehead{~~\textit{Fit Parameters}}
~~~P$_{(h)}$ &  - & 77  & days & Orbital Period \\
~~~$\sqrt{e} \cos\omega_{(h)} $ & -  & 0.15 & ... & Eccentricity Reparametrization \\
~~~$\sqrt{e} \sin\omega_{(h)} $ & - & -0.24 & ... & Eccentricity Reparametrization \\
~~~ M$_{(h)}$ & - & 120.3 & degrees & Mean Anomaly \\
~~~i$_{(h)}$ & - & 89.7 & degrees  & Inclination \\
~~~m$_{p,(h)}$ & - & $<$18.8  & 3$\sigma$ Mass Upper Limit \\
\sidehead{~~\textit{Derived Parameters}}
~~~ $\rho_{(h)}$ & - & 0.34 & g/cc & Bulk Density \\
~~~ e$_{(h)}$ & - & 0.002 & ... & Eccentricity \\
~~~ $\omega_{(h)}$ & - & 63 & Degrees & Argument of Periastron \\
~~~ K$_{(h)}$ & - & 0.6 & m s$^{-1}$ & RV Semi-amplitude \\
~~~ a$_{(h)}$ & - & 0.36 & AU & Semi-Major Axis \\
~~~ T$_{eq,(h)}$ & - & 460 & K & Equilibrium Temperature \\
\sidehead{\textbf{GP Hyperparameters}}
\hline
~~~$\eta_{1, HIRES}$ & 36.9$^{+4.5}_{-3.6}$  & 33.9$^{+2.2}_{-1.7}$ & m s$^{-1}$  & HIRES GP Amplitude \\
~~~$\eta_{1, APF}$ & 43.8$^{+4.3}_{-3.9}$ & 41.2$^{+2.7}_{-2.8}$ & m s$^{-1}$ & APF GP Amplitude \\
~~~$\eta_{1, HARPS-N}$ & 37.4$^{+6.2}_{-5.2}$ & 33.2$^{+3.5}_{-3.0}$ & m s$^{-1}$   & HARPS-N GP Amplitude \\
~~~$\eta_{2}$ & 13.5$^{+1.5}_{-2.3}$ & 13.9$^{+0.8}_{-0.9}$ & days   & Exponential Scale Length \\
~~~$\eta_{3}$ & 8.55$\pm$0.011 & 8.58$^{+0.05}_{-0.06}$ & days   & Periodic Term \\
~~~$\eta_{4}$ & 0.25$^{+0.04}_{-0.06}$ & 0.26$\pm$0.02 & ...  & Periodic Scale Length \\
\sidehead{\textbf{Instrumental Parameters}}
\hline
~~~$\gamma_{\rm{HIRES}}$ & 9.1$\pm$6.6 & 10.0$^{+3.6}_{-3.9}$ & m s$^{-1}$  & HIRES offset \\
~~~$\gamma_{\rm{APF}}$ & 4.2$\pm$7.0 & 3.9$\pm$3.9 & m s$^{-1}$ & APF offset \\
~~~$\gamma_{\rm{HARPS-N}}$ & 7.8$^{+7.5}_{-7.4}$ & 5.4$^{+4.3}_{-5.0}$ & m s$^{-1}$   & HARPS-N offset \\
~~~$\sigma_{\rm{HIRES}}$ & 1.1$^{+0.9}_{-0.7}$ & 2.1$^{+2.8}_{-1.7}$ & m s$^{-1}$    & Instrumental Jitter, HIRES \\
~~~$\sigma_{\rm{APF}}$ & 13$^{+4}_{-10}$ & 16.5$^{+2.3}_{-3.0}$ & m s$^{-1}$   & Instrumental Jitter, APF \\
~~~$\sigma_{\rm{HARPS-N}}$ & 4.7$^{+3.9}_{-3.3}$ & 7.9$^{+5.5}_{-5.8}$ & m s$^{-1}$    & Instrumental Jitter, HARPS-N \\
\enddata
\tablenotetext{*}{Although K is usually an observed parameter, it is computed in this analysis because our model parameterizes the planet masses directly.}
\tablenotetext{**}{ Estimated using an albedo of 0.}
\tablenotetext{$\dag$}{All of the orbital parameters presented in this table are osculating elements computed at BJD 2458680 days.}
\tablenotetext{***}{    We do not report uncertainties, as our model fits report overly-confident estimates that we consider unlikely to be accurate. We do not significantly detect planet (h), and so these values are not likely precise.}
\end{deluxetable*}

\begin{figure*}
    \centering
    \includegraphics[width=0.75\textwidth]{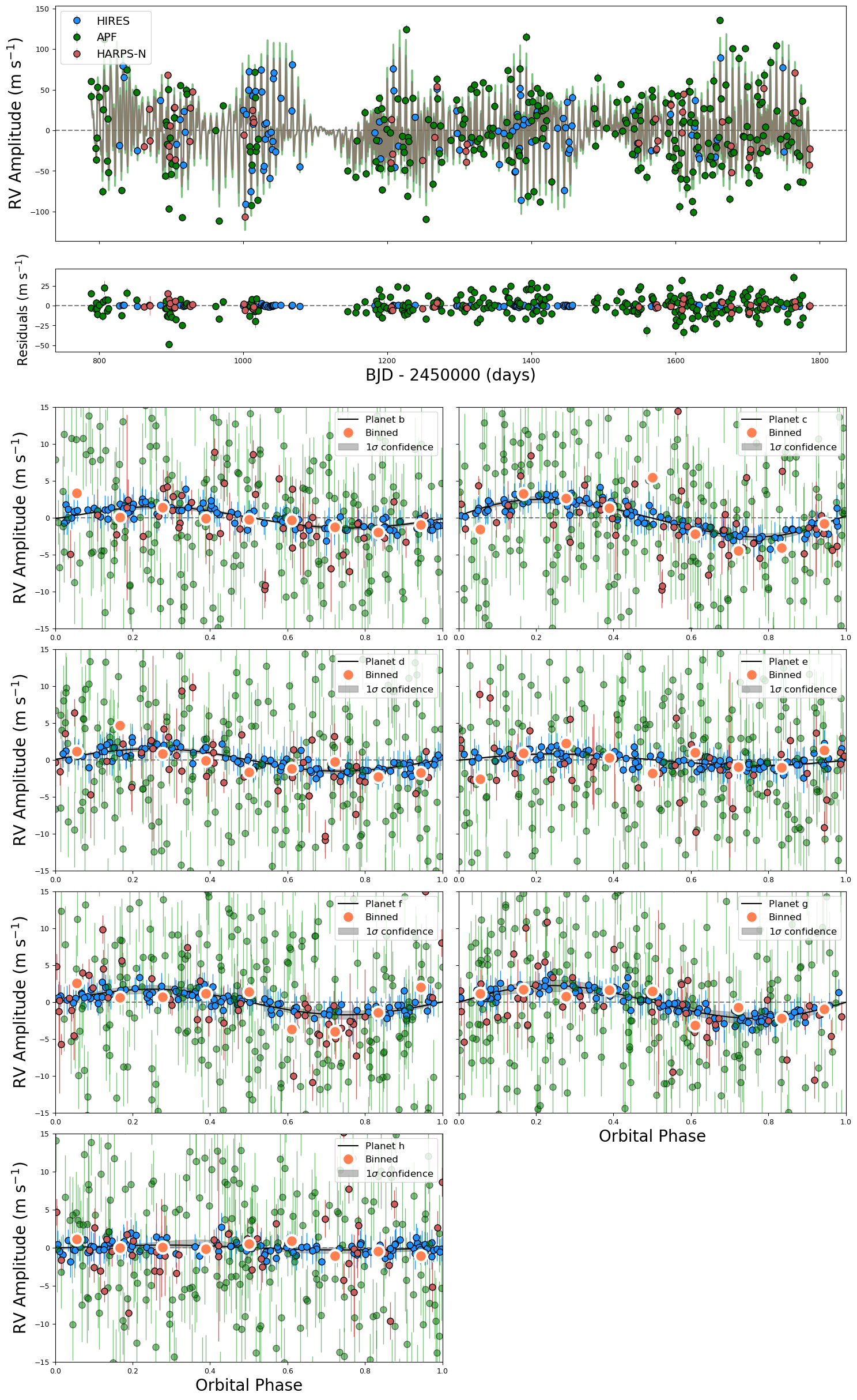}
    \caption{Top: Total RV model to TOI-1136, planets and GP. Middle: Residuals to seven planet GP fit. Bottom: Phase folds to each planet in TOI-1136 after subtracting the activity model and each other planet. RVs are adjusted for instrumental offsets. We note that APF's lower precision was not ideal for tracking planetary reflex, but helped to constrain the stellar activity. The RV data used in our analysis are available as ``data-behind-the-figure".}
    \label{fig:RV}
\end{figure*}

\begin{figure*}
    \centering
    \includegraphics[width=\textwidth]{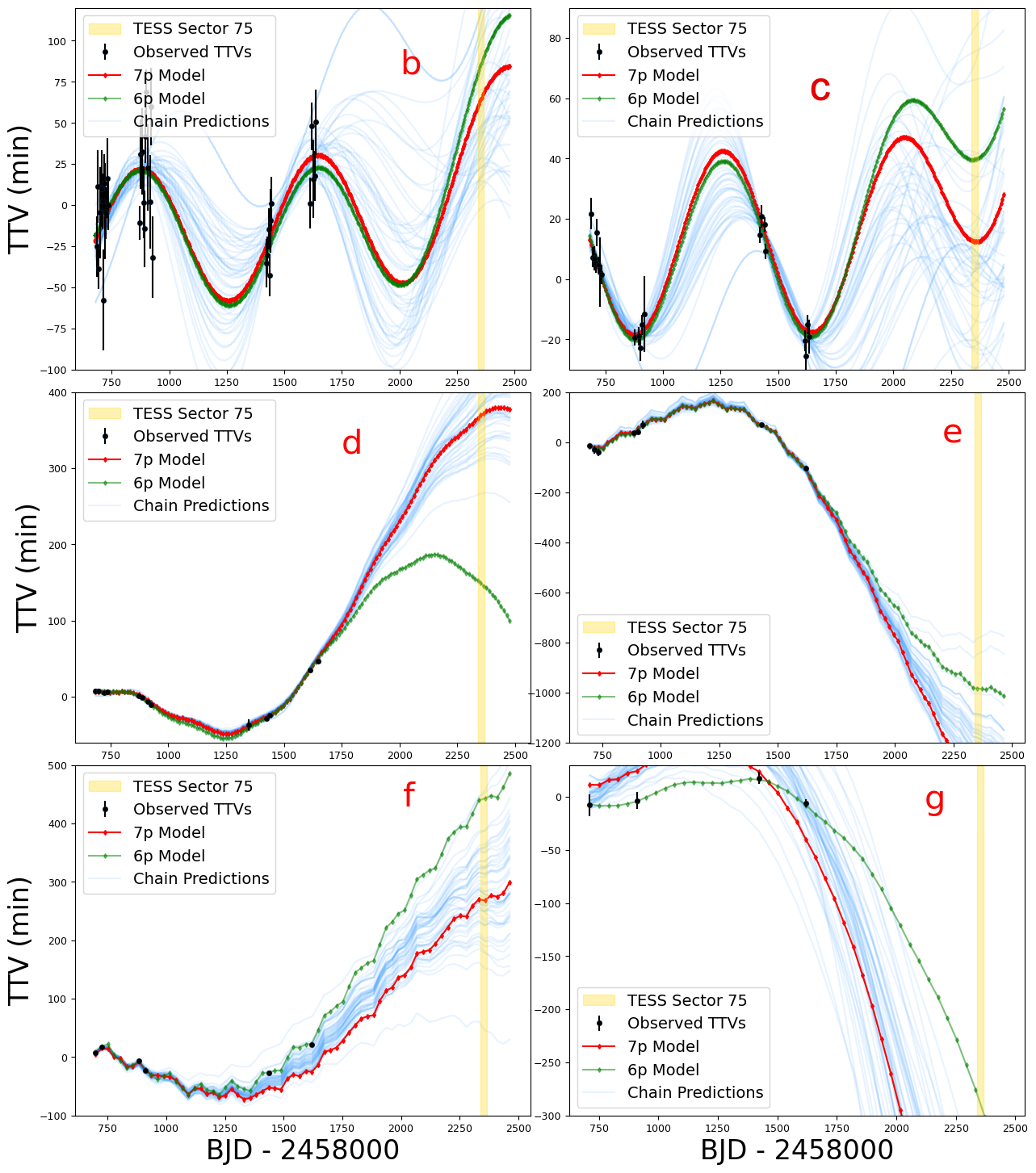}
    \caption{TTV O-C plots for each of the six inner transiting planets (b-g), from top to bottom, left to right. Red lines indicate the maximum likelihood TTV model predictions from a seven planet model, while green lines indicate a six planet model. Light blue lines indicate the final prediction of 100 randomly selected chains. We do not include a fit to the single transit of TOI-1136 (h). We also highlight TESS sector 75, where TOI-1136 will receive additional observations. The two models predict significant TTV differences during this Sector.}
    \label{fig:TTV}
\end{figure*}

\FloatBarrier
%\clearpage

\section{Discussion}\label{sec:discussion}

\subsection{Adopted Model}

We have run a total of three large analyses: a 6 planet TTV + RV fit ($\mathcal{A}$1; Table \ref{tab:posteriors}), a 7 planet TTV + RV fit ($\mathcal{A}$2; Table \ref{tab:posteriors}), and a 6 planet TTV only fit ($\mathcal{A}$3; Table \ref{tab:6pttvonly}). Not all of these results agree. For example, $\mathcal{A}$1 finds a mass of 5.6$^{+4.1}_{-3.2}$ M$_{\oplus}$ for planet g, while $\mathcal{A}$2 finds a mass of 12.0$^{+5.2}_{-3.2}$. Another example has the mass of planet d in $\mathcal{A}$2 as 5.6$^{+0.9}_{-1.0}$ M$_{\oplus}$ and in $\mathcal{A}$3 as 9.4$\pm$1.2 M$_{\oplus}$. We feel that it is worthwhile to include all of these results, especially to emphasize how differences in assumption can change model results significantly. We also feel it is best to select a single result as the primary focus of the discussion, and to choose an adopted model.

Going forward, we will mainly talk about the 6 planet TTV + RV model, $\mathcal{A}$1, and we choose this as our adopted model. We choose this over $\mathcal{A}$3 because it utilizes all of the data we have on hand, and because a TTV+RV analysis should be less susceptible to certain biases in mass measurement \citep{steffen16,mills17}. Additionally, as remarked later, the TTV+RV+GP analysis is unique and interesting for such a high multiplicity system, and selecting this model further differentiates from the detailed analysis of a 6 planet TTV-only model already carried out in D23. We reject $\mathcal{A}$2 as our preferred model because of the non-detection of planet (h) in the model, and the unreliable estimates of the planet's parameters.

\subsection{A Seven Planet System?}

While we identify a statistically significant transit in \S \ref{h_radius} that may correspond to a seventh planet, we do not significantly detect a mass for TOI-1136 (h) in our 7 planet model. Consequently, while the single transit is evidence for an additional planet in the system, we cannot confidently report its orbital period or mass. Thus, we will call this a candidate planet for the remainder of the discussion.

We spend the next sections frequently comparing TOI-1136 to the highest multiplicity exoplanet systems. We feel the comparison is appropriate because of the serious possibility that a seventh planet exists in TOI-1136, but we emphasize its status as a candidate. To reflect this nature, we will refer to the candidate at TOI-1136 (h) in various plots.

\subsection{Unique High Multiplicity Architecture}

The TOI-1136 system currently stands as a particularly unique planetary system. It is among the highest multiplicity exoplanet systems known, tied with TRAPPIST-1 \citep[7 known transiting planets;][]{gillon16,agol21} if we include the candidate planet, and just below Kepler-90 \citep[8 known transiting planets;][]{cabrera14, shallue18} and the solar system. None of these systems are alike beyond multiplicity, and TOI-1136 continues to buck the trend of similarity.

TRAPPIST-1 is an ultra-cool M dwarf with a compact architecture of planets. The planets are all terrestrial in size ($R_{p} < 1.2$ R$_{\oplus}$) and are all on short orbital periods, close to their host star ($P_{orb} < 19$ days). While the TRAPPIST-1 system has multiple potential habitable zone planets \citep{kopparapu13}, making it independently interesting, their small radii suggest that they may only have small atmospheres, and their study via transmission spectroscopy may be impossible. Already, analyses of the atmospheres of TRAPPIST-1 b and c are consistent with a no-atmosphere model \citep{greene23,ih23,lincowski23}, though \cite{krissansen23} maintain that the outer planets are still likely to have at least a small atmosphere.

Kepler-90 orbits a slightly evolved, early G dwarf, and has several longer period transiting planets. Unlike TOI-1136, Kepler-90 follows a fairly clear demarcation, with smaller, super Earth and sub-Neptune planets on shorter orbital periods, and larger gas giants on exterior orbits.

\begin{figure*}
    \centering
    \includegraphics[width=\textwidth]{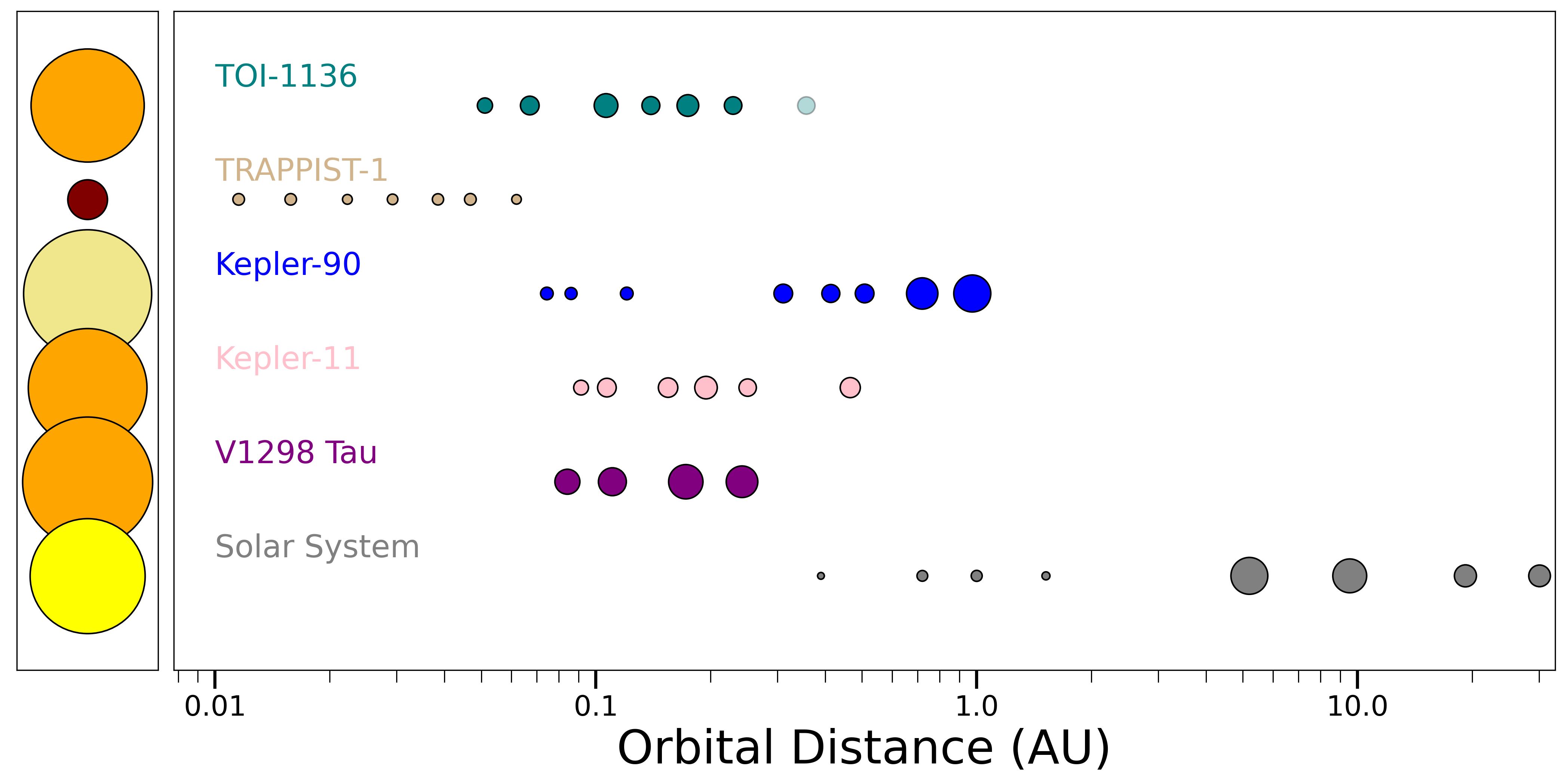}
    \caption{We highlight the disparate architectures of the highest-known multiplicity planetary systems, as well as a few systems similar to TOI-1136. We highlight that the candidate seventh planet in TOI-1136 does not have a confidently detected orbital distance. Planet and stellar radii are scaled for comparison to other systems, though we emphasize that the planet-star size is not to scale. None of the systems exhibits a clear analog to any of the others, and all have the potential for very interesting, future study.}
    \label{fig:architectures}
\end{figure*}

TOI-1136 consists entirely of sub-Neptune sized planets, likely none of them terrestrial. Further, none are large enough to call gas giants, either, and the planet sizes do not follow any clear sequence or demarcation, with the largest planet third from the star. We highlight the architectural differences in Figure \ref{fig:architectures}. TOI-1136's youth is yet another distinguishing feature that adds to the system's value.

Kepler-11 is perhaps the most similar \textit{Kepler} system to TOI-1136, with six transiting planets orbiting a G dwarf \citep{lissauer13}. Additionally, its six planets are all similar in size (R$_{p}$ = 1.8 - 4.2 R$_{\oplus}$) and density ($\rho$ = 0.58 - 1.4 g/cc) to TOI-1136. As with many \textit{Kepler} systems, the low brightness of Kepler-11 makes RV observations difficult, making any combined analysis like that of TOI-1136 more challenging, though a TTV + RV analysis was done in \cite{weiss_dissertation}. This system is also likely not young, making some science cases less promising.

We finally compare TOI-1136 to V1298 Tau, a very young system with four transiting exoplanets \citep{suarez22}. The system is even brighter than TOI-1136, and the star is even younger ($\sim$ 20 Myr). V1298 Tau's RVs are much more contaminated with stellar activity than even TOI-1136, making its study very challenging \citep{blunt23}. The system does not appear to exhibit TTVs, however, making its mass extraction much more difficult than TOI-1136.

While many systems exhibit many of the attractive features present in TOI-1136 (multiplicity, youth, TSM), few others have all in the right combination to allow for precise mass measurement, as is possible with TOI-1136. 

\subsection{Resonance Gives Insight into Formation}

D23 performed an extensive analysis of TOI-1136, especially considering the resonant orbital properties of neighboring planets, and the overall dynamical stability of the system. Unlike most \textit{Kepler} systems, the orbital periods of TOI-1136 do not deviate from resonance by more than 1$\%$. Particularly strange is the existence of a second-order resonance between planets e and f, which is rare and usually unstable. The youth of TOI-1136 ($\sim$ 700 Myr) suggests the possibility that TOI-1136 is a young precursor to more mature \textit{Kepler} systems, and perhaps suggests that higher order resonances are more common than observed, but become unstable on shorter timescales.

This unique characteristic of TOI-1136 allows us to make more sophisticated guesses about the system's formation and evolution. The system can in many ways be likened to a snapshot of a younger \textit{Kepler} system. Our constraints on eccentricity and argument of periastron, in particular, may shed light beyond the analysis in D23. When experiencing Type-I migration, planets will often form far from the star, and move inward via mutual interactions and disk torque. Theory suggests that such migration results in opposite arguments of periastron to minimize mutual interactions \citep{batygin13}. We include a similar figure to Figure 19 in D23 (Figure \ref{fig:anti_aligned}). With the exception of planets c and d, and the candidate, posteriors are highly suggestive of Type-I migration. Future atmospheric studies would likely help confirm if indeed the planets in TOI-1136 migrated inward, possibly from the beyond the ice line.

\subsection{Improved Mass Precisions?}

TKS began observations of TOI-1136 well before its true multiplicity was known, and before any significant transit time variations were detected. As our knowledge of the system evolved, the large number of RVs acquired for the system became less obviously useful: with the high mass precisions measured in D23 for the 6 planets using TTVs alone, the RVs seemed unlikely to improve our mass constraints by a great deal. RV-only fits were hindered by several challenges, preventing significant detections of most of the planets. Mainly, the stellar variability amplitude was many times larger than the expected RV semi-amplitudes, and the stellar rotation period was close to several of the planet orbital periods. With the relatively poor cadence of RV data (compared with photometry), disentangling Keplerian signals from stellar variability became very hard to do with confidence.

Our adopted model generally extracts mass precisions and values consistent with D23. Planet's c, d, and g see slightly improved mass precisions, while the others see slightly worse. Our 7 planet model and our TTV-only models, however, see generally much more precise masses, and in some cases masses quite distinct from D23. It may be that including a seventh planet improves the model significantly, though we consider this unlikely considering its insignificant detection and possibly incorrect orbital period. Additionally, the inclusion of RVs may not be the only contribution to our adopted model's differing posterior parameters. Figure \ref{fig:ttv_ttvrv} shows our results, compared with a TTV-only model run using \texttt{TTVFast} and \texttt{emcee}. It is clear that, especially for the inner planets, the fits which include RVs are not more precise. They are, in fact, typically less well constrained than a TTV-only fit. This suggests that the resulting differences are more likely caused by a different N-body integrator, sampler, or both. D23 utilized \texttt{JAX} \citep{bradbury18} for N-body integration, and a No U-Turn Sampler \citep[NUTS;][]{betancourt17} for inference. We utilized \texttt{TTVFast} for N-body integration and \texttt{emcee} for sampling. 

An analogous situation may be the mass measurements of TRAPPIST-1 in \cite{wang17}, which utilized \texttt{TTVFast} and \texttt{emcee}, that were later rectified in \cite{agol21} using a NUTS sampler. The situation is not perfectly analogous, however, as the masses reported in \cite{wang17} were highly discrepant with those in \cite{agol21}, which is not the case between our mass estimates here, and the values reported in D23. Additionally, the uncertainties reported in \cite{wang17} were much larger than the values reported in \cite{agol21}, which is only the case for three planets in our adopted model, and the difference is not large. Additionally, we know of at least two multi-planet systems with a TTV + RV analysis that utilize \texttt{TTVFast} and \texttt{emcee} in conjunction \citep[Kepler-11; WASP-47;][]{weiss_dissertation,almenara16}, suggesting that the combination is not necessarily unreliable. Convergence and other sanity checks do not suggest issues during inference, and so we report our results here with a caution that the 7 planet fit and the TTV-only fit have discrepancies with D23, and we are not entirely certain of the cause. Our adopted model, however, is generally consistent.

The high amounts of correlated noise in the RVs are the most likely culprit lowering the precision of our TTV + RV models. Despite this, we include them in our model for a number of reasons. A 6 planet TTV + RV model is generally more consistent with D23. Including RVs also prevents our results from biasing towards the known systematic differences between TTV masses and RV masses \citep{steffen16,mills17}. Further, the additional complication added to the analysis by utilizing a TTV + RV + GP model, we feel, is a useful case study for the field, regardless of the result.  

Few exoplanet studies are capable of utilizing both RVs and dynamical TTVs, and those systems that are amenable typically have lower multiplicity. Many high-multiplicity systems are analyzed by their TTVs alone \citep[e.g.][]{lissauer13,agol21}, or their RVs alone \citep[i.e.][]{motalebi15,feng17,santerne19,lubin22,turtelboom22}, and lower-multiplicity systems have seen combined analyses \citep[i.e.][]{weiss16, almenara16, weiss17}. The only other high multiplicity system ($>$ 5 planets) for which RVs and TTVs are jointly modeled is Kepler-11 \citep{weiss_dissertation}. \cite{weiss_dissertation} found that including RVs did not improve mass measurements of Kepler-11 appreciably compared to TTV-only fits, though a comparison is imperfect as this analysis only utilized 27 RVs, in contrast to the 410 RVs used in our analysis of TOI-1136. Our analysis appears to be the first for which an N-body forward model with Gaussian process is jointly fit to the TTVs and  RVs.   A full photodynamical analysis of the photometry jointly with the RVs, including a model for stellar activity in both the photometry and the RVs, might further improve the planet mass and orbit determinations, but such an effort is enormously computationally costly and is beyond the scope of this paper.

\subsection{Prospects for Atmospheric Studies}

\begin{figure*}
    \centering
    \includegraphics[width=\textwidth]{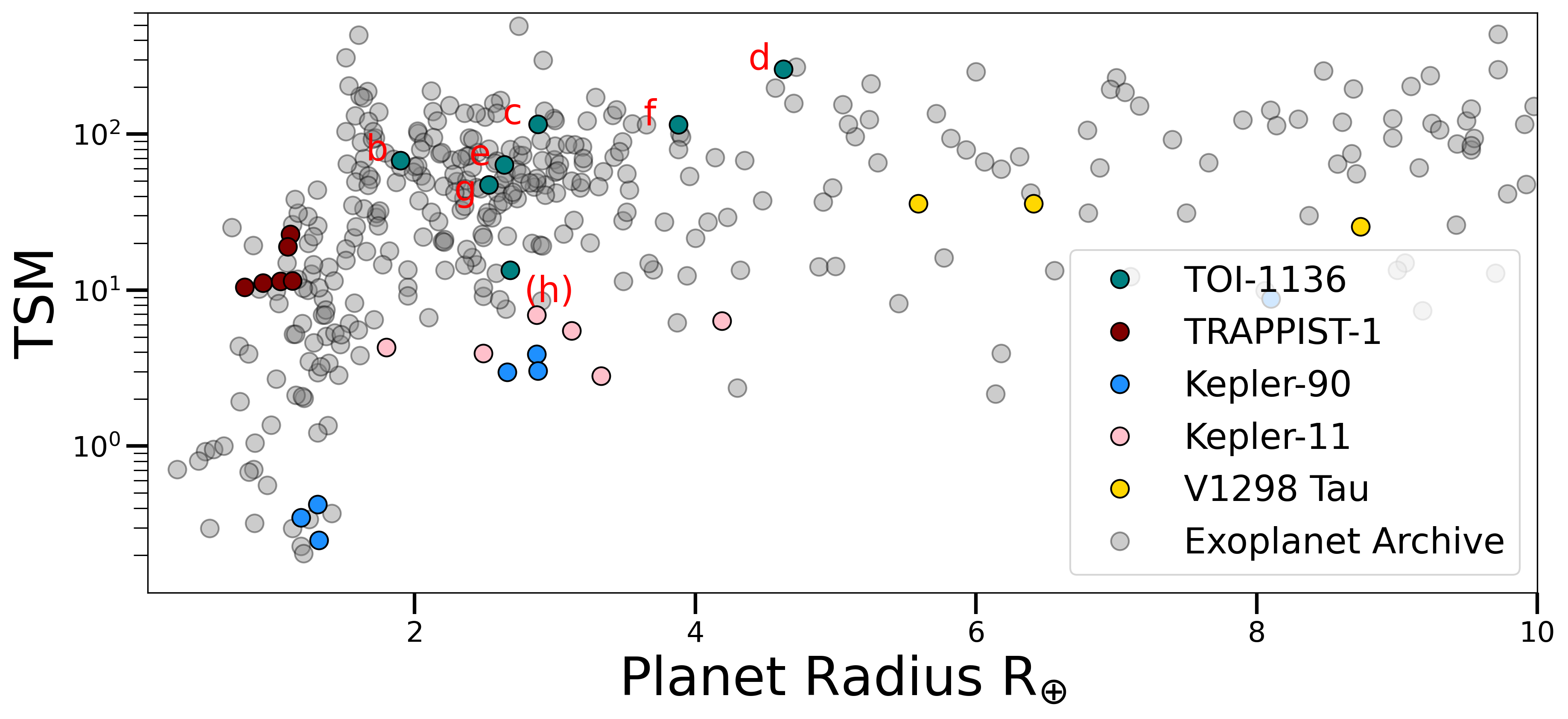}
    \caption{TSM versus planetary radius of known exoplanets, taken from the \texttt{Exoplanet Archive} on 25 October 2023. We also highlight TRAPPIST-1, Kepler-90, Kepler-11, and V1298 Tau, the systems we discussed as most relevant for comparison with TOI-1136. None of these other systems have TSM values as high as TOI-1136. While several planets in TOI-1136 have only average TSM values, planets c and d are very good for follow up. We emphasize that the probable existence of an atmosphere on all planets in the TOI-1136 system inflates the system's TSM values, and the system would be less useful for a study focused on terrestrial planets.}
    \label{fig:TSM}
\end{figure*}

The potential for future atmospheric studies is a significant portion of TOI-1136's value to the scientific community. The bulk densities of all six transiting planets, and the candidate seventh, are consistent with appreciable atmospheric envelopes, suggesting that atmospheric features may be detected on all seven planets. The transmission spectroscopy metric \citep[TSM; ][]{kempton18} is a useful metric for assessing the value of transmission spectroscopy for a variety of planet regimes. Planets b-g have TSM values of 68, 116, 260, 64, 115, and 47, respectively. These values are estimated assuming an albedo of 0, as is often done \citep[e.g.][]{beard22}. TOI-1136 c and TOI-1136 d both rank higher than the follow-up cutoffs suggested in \cite{kempton18}, and planet d is particularly good, ranking in the second quartile of large planets. A comparison of TOI-1136 TSMs with other published exoplanets is shown in Figure \ref{fig:TSM}.

The true value of studying TOI-1136 via transmission spectroscopy comes from a combination of its multiplicity and its youth. Multiplicity allows for comparative exoplanetology between planets in the same system. This is advantageous because the formation environment of a planetary system is a considerable source of uncertainty, and studying multiple planets in the same system allows for the removal of this uncertainty \citep{owen20}. Comparing the different environments and atmospheres between the planets of TOI-1136 would provide a great deal of information about the processes that formed the planets in the system, especially their dependence on non-stellar parameters. Is the composition of all the planets the same? If they differ, does it depend on orbital period or eccentricity? Have interior planets been noticeably depleted of volatiles by XUV sculpting? 

The youth of TOI-1136 suggests that the system is likely still evolving. Some studies suggest atmospheric stripping may occur on Myr timescales \citep{sanzforcada11}, while others suggest that it continues into the Gyr regime \citep{berger20_a}. For example, the high insolation received by many of the planets in the system might plausibly strip the atmospheres of the inner planets, if it has not done so already. The largest planet in the system, however, is third closest to the star, in contrast to the typical architectures seen in multi-planet systems \citep{lissauer11}. This suggests that atmospheric stripping may be ongoing in this system. Preliminary atmospheric observations of TOI-1136 d detect H$\alpha$ absorption, a possible sign of atmospheric stripping (Orell-Miquel et al., in prep). Furthermore, the stellar type of TOI-1136 is very similar in parameter space to the Sun, which offers particularly strong motivation for additional study. While the planetary environment does not appear at all similar to the Solar system, the evolution of TOI-1136 could inform predictions about the evolution of our own home.

The youth of TOI-1136, while adding to the potential scientific interest of transmission spectroscopy, might also hinder spectral models. Spectral contamination, however, most strongly hinders low-resolution spectroscopy of late type stars, and earlier type stars mainly see contamination in optical wavebands, which is less of an issue for JWST \citep{frederic22}. Additionally, such contamination can be mitigated by high resolution spectroscopy, which we have in abundance for TOI-1136.

\begin{figure*}
    \centering
    \includegraphics[width=\textwidth]{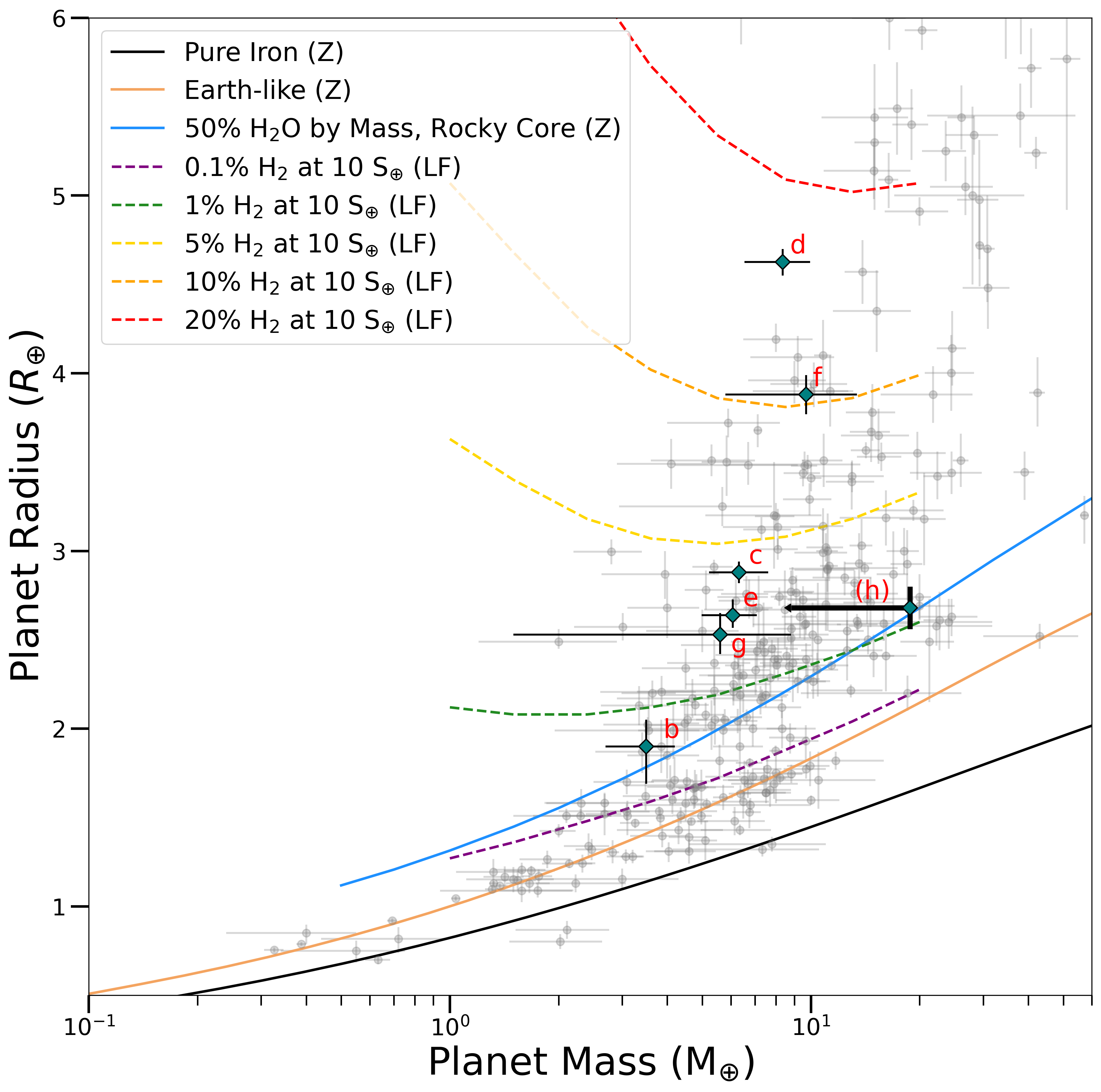}
    \caption{Mass-radius diagrams of known exoplanets taken from the \texttt{NASA Exoplanet Archive} on 12 May 2023 in gray, with the planets in TOI-1136 highlighted. We include only exoplanets with better than 2$\sigma$ mass precision. We include composition profiles taken from \cite{zeng19} for rock, water, and iron compositions, indicated by a solid line in the figure, and a (Z) in the legend. We include H$_{2}$ envelopes of different percentages taken from \cite{lopez14}, as the \cite{zeng19} profiles may not be as accurate in the regime of large gaseous envelopes \citep{rogers23}. These are indicated by dashed lines, and are notated with a (LF) in the plot legend. A wide variety of compositions might explain the bulk density of the planets in TOI-1136, and planet b in particular might have either a small volatile envelope, or could be consistent with a ``water world". We place TOI-1136(h) at its 3$\sigma$ upper limit, and use a downard arrow to indicate our uncertainty in its mass.}
    \label{fig:MR}
\end{figure*}

Comparing the various planets of TOI-1136 with other known planets can be highly suggestive of their compositions. We put the 6 known planets, and the candidate planet, on a mass-radius diagram in Figure \ref{fig:MR}. 

The possible compositions of the planets in TOI-1136 depend strongly on the insolation. Planets in TOI-1136 are hot, with insolations of 365, 213, 84, 49, 31, and 18 S$_{\oplus}$. \cite{lopez14} only estimate compositions curves for a limited number of insolations, the closest being 10 S$_{\oplus}$ and 1000 S$_{\oplus}$. Despite these caveats, the placement of the planets in mass-radius space suggests a wide variety of possible compositions for every planet in the system, and follow-up study with JWST would likely reveal a great deal about the chemicals in the atmospheres of these plaents.

Planet b is in the radius gap \citep{fulton17}, and might realistically have a terrestrial or gaseous composition. Figure \ref{fig:MR} suggests a large envelope of water vapor may be the best description of TOI-1136 b's atmosphere, though a variety of volatile envelopes could presumably describe the planet as well. D23 made a strong case that TOI-1136 has experienced Type-I migration, which makes planet b an excellent water world candidate. The resonance of the system suggests that planets likely migrated inward, which is one of the primary ways an exoplanet so close to its host star might still contain significant amounts of water. Our new constraints on the argument of periastron of the planets in the system further suggest Type-I migration (Figure \ref{fig:anti_aligned}), as neighboring planets are expected to have anti-aligned arguments of periastron \citep{batygin13}.

Planet c, d, e, and f, on the other hand, seem consistent with a large gaseous envelope of H$_{2}$ or some other volatiles. Even among these planets, compositions vary appreciably, with planets d and f likely containing larger envelopes of H$_{2}$, while planets c and e are notably less ``puffy". Stellar winds may have stripped some of their atmospheres, but it remains a mystery as to why planet e would experience such stripping at an increased rate as compared to planets d and f.

TOI-1136 (h) does not have stringent mass or orbital period measurements, and we cannot say much about its potential composition, except that it likely contains a gaseous envelope of some kind. Future studies that confirm or refute the planetary nature of this signal could shed a great deal of additional light on its theoretical composition.

\subsection{Bridging the Radius Gap}

A dearth of exoplanets with radii between 1.5 R$_{\oplus}$ and 2.0 R$_{\oplus}$ was first identified in \cite{fulton17}, and since has been of great interest to the exoplanet community. This line seems to demarcate terrestrial planets from more gaseous sub-Neptunes, and exoplanets within the gap, in particular, could be subject to either composition. Studying systems with planets on either side of the radius gap can give special insight to the formation, and a number of such studies have been carried out in the literature \citep{crossfield19,nowak20}. TOI-1136 is an extremely useful system to include in such studies, as it has 1 planet within the radius gap (b), 5 planets above the radius gap (c, d, e, f, g), and a candidate planet above the radius gap (h). A great deal of information might be gleaned from a follow-up study examining each planet's expected role in such a configuration, though it is beyond the scope of our analysis here.

\subsection{Future Work}

We expect TOI-1136 to receive continued observations and scientific interest. Many of the most attractive features of the system, such as its amenability to transmission spectroscopy, are due to the possibility of future observations. TOI-1136 makes for an extremely compelling target for \textsf{JWST}.

Future RV observations might better constrain the mass of the system, though future TESS observations are likely to be more fruitful. TTVs seem to contribute a great deal to the mass precision of the system's exoplanets, and more transits should only further refine our knowledge. Other parameters, such as radius, orbital period, and time of conjunction will see improvements with more TESS observations. Observing additional transits of the candidate planet would be the best way to confirm its planetary nature. Fortunately, TESS will be re-observing TOI-1136 in Sector 75, which starts on 30 January 2024. 

This system is particularly interesting in the context of the observed discrepancy between TTV and RV measured exoplanet masses \citep{steffen16,mills17}. Very few exoplanet systems with TTV masses are also amenable to RV follow-up ($<7$; \texttt{NASA Exoplanet Archive}). Recovering significant mass measurements with RVs alone, while preventing GP overfitting, would be a challenging task, probably requiring many more observations, but could potentially shine light on this discrepancy. It might additionally alleviate concerns raised in \S \ref{sec:cross validation} about model overfitting, as it would be interesting to ensure the two methods are consistent.

\section{Summary}\label{sec:summary}

We utilize a combination of TTVs, RVs, and a GP to measure the mass of the six-planet system TOI-1136, and place constraints on the orbital properties of a potential seventh planet. This detailed analysis will inform future studies of TOI-1136, as the system is a top candidate for transmission spectroscopy, and is a huge source of potential information about planetary formation.

\section{Acknowledgements}

The primary TTV + RV analysis was carried out by C.B. with help from F.D. and P.R. Frequency analysis and additional figures were carried out with help from R.H. and J.L. J.A.M helped improve the atmospheric models used in our discussion section, and S.B. and R.A.R. helped contributed to cross-validation analysis. G.N. helped facilitate communication between the TKS and the HARPS-N collaborations. The TKS survey was designed and managed by the PIs N.M.B., I.C., C.D., B.F., A.W.H., D.H., H.I., S.R.K, E.A.P., A.R., and L.M.W. The TKS target pool was developed with help from T.M., M.L.H., J.L., C.B., A.W.M., J.A.M., and N.S. Final survey targets were developed by A.C., T.F., J.V.Z., J.A.M., C.B., and R.A.R. HIRES observations were acquired with help from C.L.B., D.T., A.W.M., M.M., A.C., J.A.M., C.B., S.G., S.B., F.D., J.V.Z, J.L., M.R., L.W., A.P., P.D., R.A.R., and A.B. HARPS-N day-to-day operations were executed with help from G.N., I.C., F.M., J.O., R.L., and E.P. Observations were obtained with help from J.K., R.B., G.M., M.O., and K.K.

%KOA

The data presented herein were obtained at the W. M. Keck Observatory, which is operated as a scientific partnership among the California Institute of Technology, the University of California and the National Aeronautics and Space Administration. The Observatory was made possible by the generous financial support of the W. M. Keck Foundation.

Additionally, the authors wish to recognize and acknowledge the very significant cultural role and reverence that the summit of Maunakea has always had within the indigenous Hawaiian community. We are most fortunate to have the opportunity to conduct observations from this mountain.

%Anonymous Referee

We wish to thank the anonymous referee for their diligent and detailed feedback. We feel that they have improved the paper significantly.

%Gaia DR3

This work has made use of data from the European Space Agency (ESA) mission
{\it Gaia} (\url{https://www.cosmos.esa.int/gaia}), processed by the {\it Gaia}
Data Processing and Analysis Consortium (DPAC,
\url{https://www.cosmos.esa.int/web/gaia/dpac/consortium}). Funding for the DPAC
has been provided by national institutions, in particular the institutions
participating in the {\it Gaia} Multilateral Agreement.

%TESS Acknowledgement
This paper includes data collected by the TESS mission. Funding for the TESS mission is provided by the NASA's Science Mission Directorate. TESS data used during the analysis of this system are available at the permananet DOI link \url{https://archive.stsci.edu/doi/resolve/resolve.html?doi=10.17909/t9-st5g-3177}.

%UCI HPC acknowledgement
This work utilized the infrastructure for high-performance and high-throughput computing, research data storage and analysis, and scientific software tool integration built, operated, and updated by the Research Cyberinfrastructure Center (RCIC) at the University of California, Irvine (UCI). The RCIC provides cluster-based systems, application software, and scalable storage to directly support the UCI research community. https://rcic.uci.edu

%Grants and funding

%TESS guest investigator
This work was partially supported by NASA grant 80NSSC22K0120 to support Guest Investigator programs for TESS Cycle 4. 

%FINESST
This work was partially support by the Future Investigators in NASA Earth and Space Science and Technology (FINESST) program Grant No. 80NSSC22K1754.

% NSF GRFP
J.M.A.M. is supported by the National Science Foundation (NSF) Graduate Research Fellowship Program under Grant No. DGE-1842400.

T.F. acknowledges support from the University of California President's Postdoctoral Fellowship Program.

G.N. thanks for the research funding from the Ministry of Education and Science program the "Excellence Initiative - Research University" conducted at the Centre of Excellence in Astrophysics and Astrochemistry of the Nicolaus Copernicus University in Toru\'n, Poland.%Grzegorz Nowak

R.L. acknowledges funding from University of La Laguna through the Margarita Salas Fellowship from the Spanish Ministry of Universities ref. UNI/551/2021-May 26, and under the EU Next Generation funds. 

G. M. acknowledges funding from the Ariel Postdoctoral Fellowship program of the Swedish National Space Agency (SNSA)

M.R. thanks the Heising-Simons Foundation for their generous support.

L.M.W. acknowledges support from the NASA-Keck Key Strategic Mission Support program (grant no. 80NSSC19K1475) and the NASA Exoplanet Research Program (grant no. 80NSSC23K0269)

This work is partly supported by JSPS KAKENHI Grant Number JP21K13955.

J.~K. gratefully acknowledges the support of the Swedish National Space Agency (SNSA; DNR 2020-00104) and of the Swedish Research Council (VR: Etableringsbidrag 2017-04945).

\facilities{\gaia{}, Keck (HIRES), APF, TNG (HARPS-N), \tess{}, Exoplanet Archive, UCI RCIC}
\software{
\texttt{ArviZ} \citep{exoplanet:arviz},
\texttt{emcee} \citep{foremanmackey13},
\texttt{exoplanet} \citep{Foreman-Mackey2021,exoplanet:zenodo},
\texttt{ipython} \citep{ipython07},
\texttt{lightkurve} \citep{lightkurve18},
\texttt{LMFit} \citep{newville14},
\texttt{matplotlib} \citep{Hunter07},
\texttt{numpy} \citep{harris20},
\texttt{pandas} \citep{reback20,mckinney-proc-scipy-2010},
\texttt{PyMC3} \citep{exoplanet:pymc3},
\texttt{RadVel}\citep{fulton18},
\texttt{SciPy} \citep{2020SciPy-NMeth},
\texttt{theano} \citep{exoplanet:theano},
\texttt{TTVFast} \citep{deck14}
}

\bibliography{bibliography}

\appendix

\renewcommand{\thefigure}{A\arabic{figure}}
\setcounter{figure}{0}
\renewcommand{\thetable}{A\arabic{table}}
\setcounter{table}{0}

We include a number of additional tables and figures in the appendix that may be of interest.

\FloatBarrier

\begin{figure*}
    \centering
    \includegraphics[width=\textwidth]{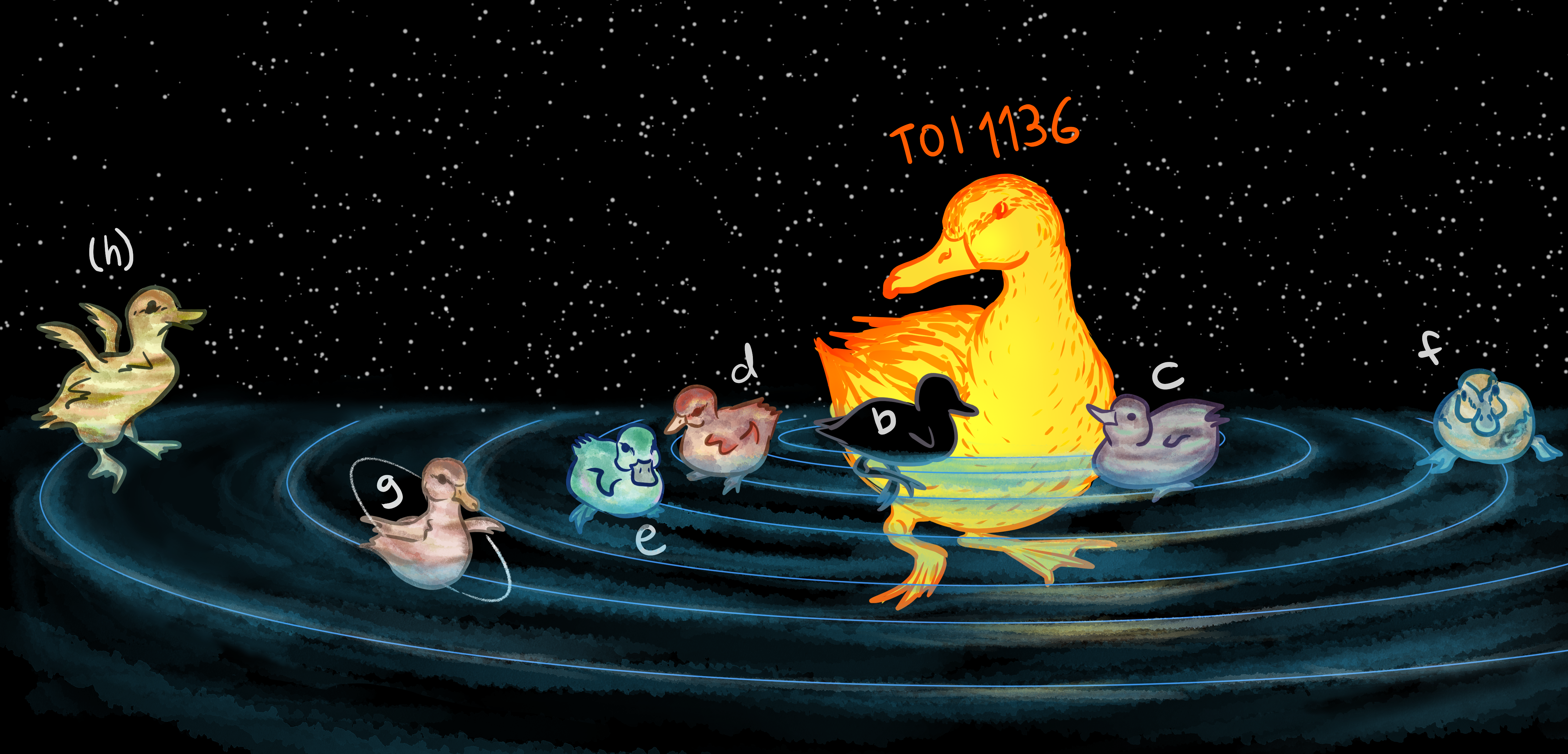}
    \caption{Here we present an amusing rendition of the TOI-1136 system if each body in the system were a duck or duckling, created by co-author Rae Holcomb. We encourage any future promotions of work associated with TOI-1136 to use this graphic at their leisure.}
    \label{fig:ducks}
\end{figure*}

\begin{figure*}
    \centering
    \includegraphics[width=\textwidth]{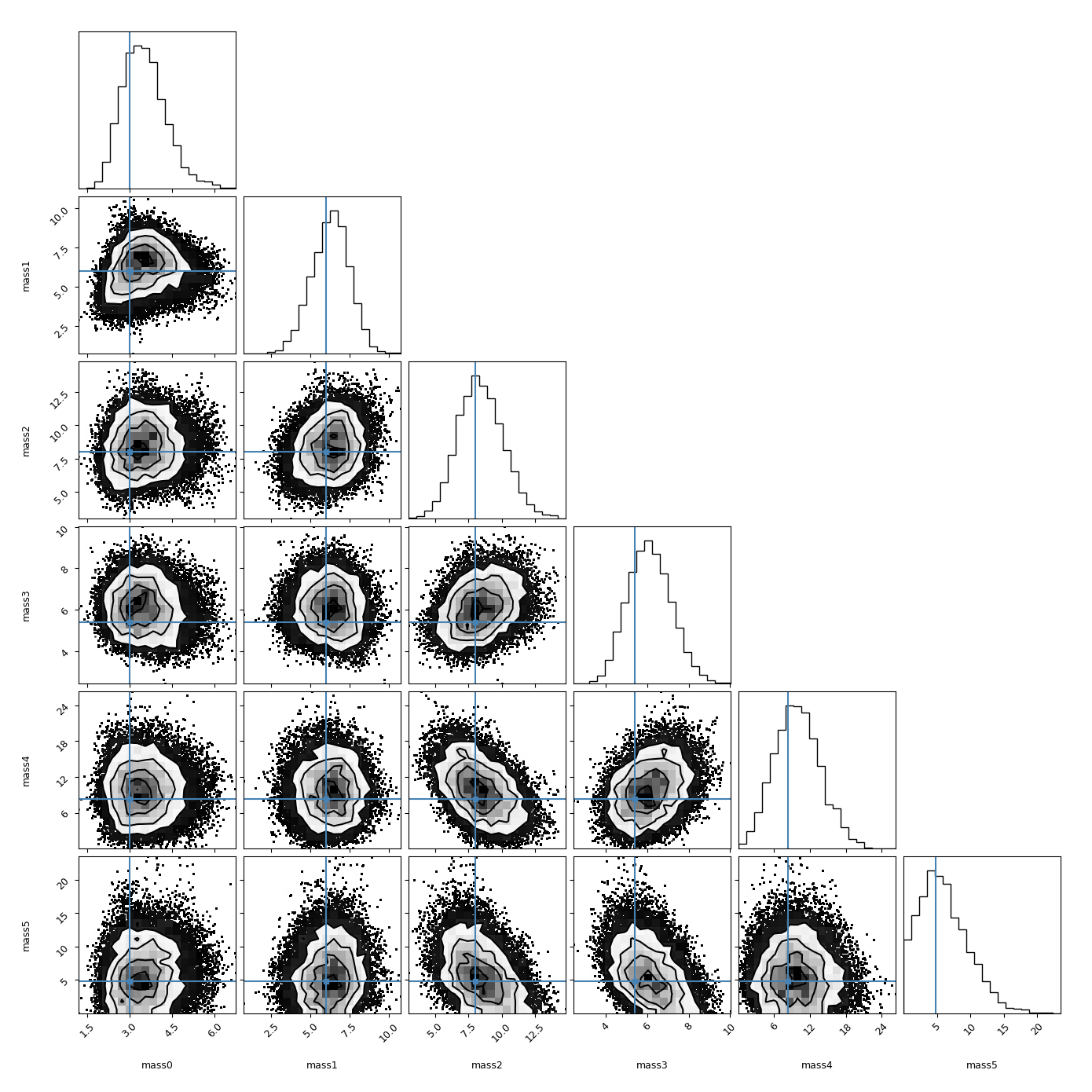}
    \caption{Corner plot highlighting the mass fits to each planet in our adopted model. Blue lines indicate the value reported in D23.}
    \label{fig:TTV-OC}
\end{figure*}

\begin{figure*}
    \centering
    \includegraphics[width=\textwidth]{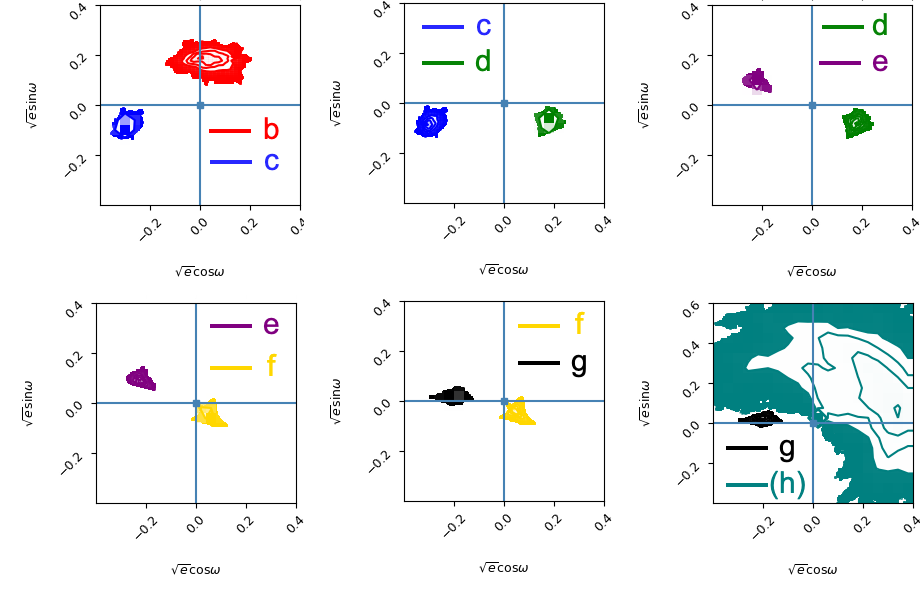}
    \caption{We include comparisons of the $\sqrt{e}\cos\omega$ and $\sqrt{e}\sin\omega$ posteriors for each planet in the system. \cite{batygin13} predict that the argument of periastrons of neighboring planets should be anti-aligned when in resonance. Planets in the TOI-1136 system seem to generally follow this principle, a strong indicator that the system experienced Type-I migration.}
    \label{fig:anti_aligned}
\end{figure*}

\begin{figure*}
    \centering
    \includegraphics[width=\textwidth]{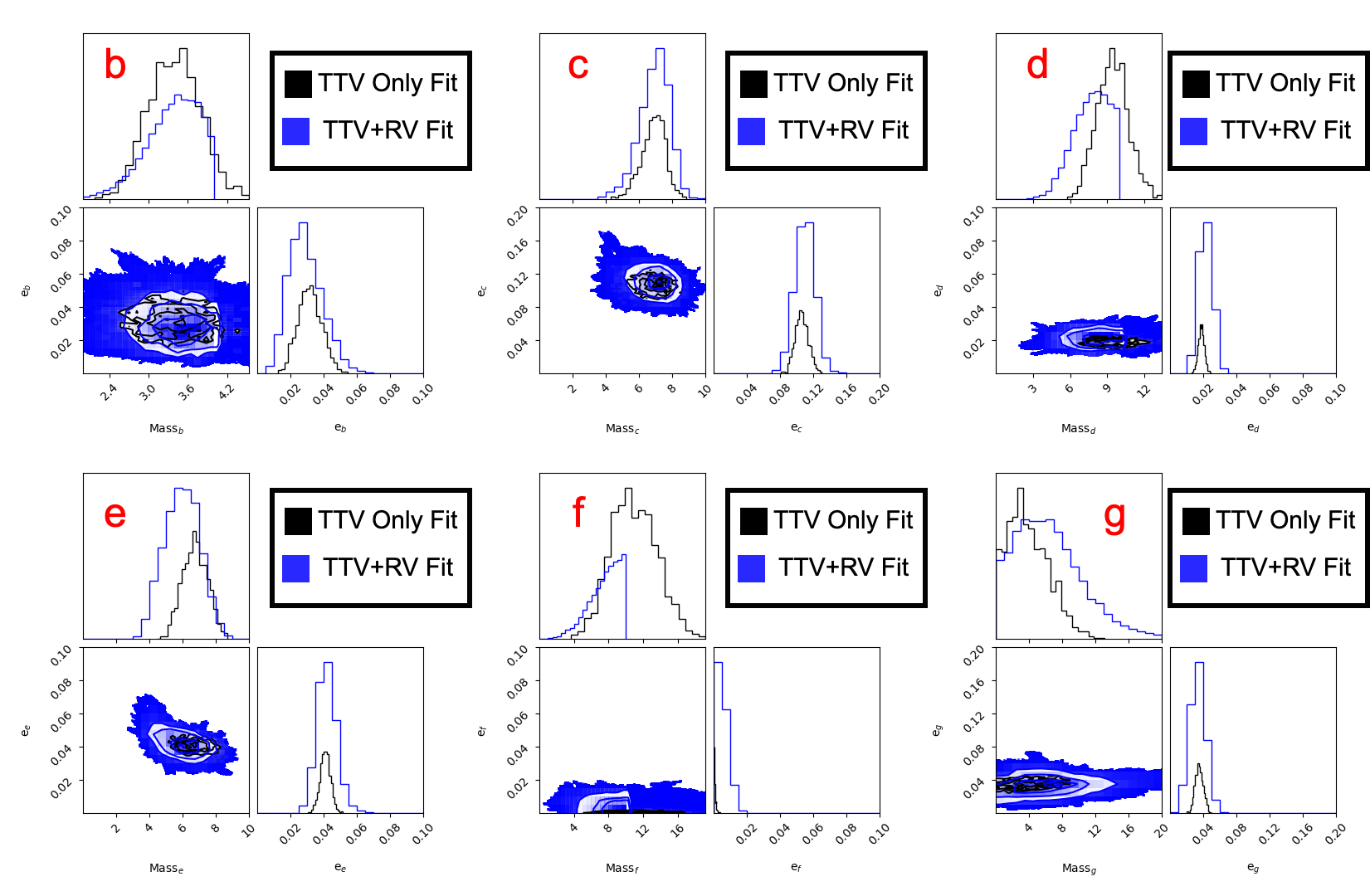}
    \caption{A comparison of the posterior samples for mass and eccentricity of a TTV-only 6 planet fit, to our TTV + RV + GP 6 planet fit. The TTV-only posteriors are more precise in most cases, indicating that improved posterior estimates in our model are not necessarily due to inclusion of RVs, but may also be sampler-dependent.}
    \label{fig:ttv_ttvrv}
\end{figure*}

\clearpage

\startlongtable
\begin{deluxetable*}{llll}
\centering
\tablecaption{TTV-Only Posteriors of TOI-1136$^{\dag}$}
\label{tab:6pttvonly}
\tablehead{\colhead{~~~Parameter Name} & 
\colhead{6p TTV Posterior}  & \colhead{Units} & \colhead{Description}}
\startdata
\sidehead{\textbf{Planet b}}
\hline
\sidehead{~~\textit{Fit Parameters}}
~~~P$_{b}$ & 4.1728$\pm$0.0002  &  days & Orbital Period \\
~~~$\sqrt{e} \cos\omega_{b} $ & 0.15$\pm$0.02 &   ... & Eccentricity Reparametrization \\
~~~$\sqrt{e} \sin\omega_{b} $ & 0.09$^{+0.03}_{-0.02}$  &  ... & Eccentricity Reparametrization \\
~~~ M$_{b}$ & 46.7$^{+7.3}_{-6.5}$ &  degrees & Mean Anomaly \\
~~~i$_{b}$ & 86.43$\pm$0.27 &   degrees  & Inclination \\
~~~m$_{p,b}$ & 3.38$^{+0.41}_{-0.38}$ &  M$_{\oplus}$ & Planet Mass \\
\sidehead{~~\textit{Derived Parameters}}
~~~ $\rho_{b}$ &  2.71$\pm$0.84 & g/cc & Bulk Density \\
~~~ e$_{b}$ & 0.03$\pm$0.01 &  ... & Eccentricity \\
~~~ $\omega_{b}$ & 31$\pm$7 &  Degrees & Argument of Periastron \\
~~~ a$_{b}$ & 0.0511$\pm$0.0006 &  AU & Semi-Major Axis \\
~~~ T$_{eq,b}^{**}$ & 1216$\pm$10  &  K & Equilibrium Temperature \\
\sidehead{\textbf{Planet c}}
\hline
\sidehead{~~\textit{Fit Parameters}}
~~~P${_c}$ & 6.2573 $\pm$ 0.0002 &  days & Orbital Period \\
~~~$\sqrt{e} \cos\omega_{c} $ & -0.112$^{+0.009}_{-0.010}$  &  ... & Eccentricity Reparametrization \\
~~~$\sqrt{e} \sin\omega_{c} $ & -0.305 $\pm$ 0.010 &  ... & Eccentricity Reparametrization \\
~~~ M$_{c}$ & 63.4$^{+1.7}_{-1.6}$ &   degrees & Mean Anomaly \\
~~~i$_{c}$ & 89.31$^{+0.40}_{-0.50}$ &    degrees  & Inclination \\
~~~m$_{p,c}$ & 6.90$^{+0.63}_{-0.75}$ &  M$_{\oplus}$ & Planet Mass \\
\sidehead{~~\textit{Derived Parameters}}
~~~ $\rho_{c}$ & 1.58$\pm$0.19 &  g/cc & Bulk Density \\
~~~ e$_{c}$ & 0.11$\pm$0.01 & ... & Eccentricity \\
~~~ $\omega_{c}$ & 70$\pm$2 &  Degrees & Argument of Periastron \\
~~~ a$_{c}$ & 0.0669$\pm$0.0005 &  AU & Semi-Major Axis \\
~~~ T$_{eq,c}$ & 1062$\pm$7 &  K & Equilibrium Temperature \\
\sidehead{\textbf{Planet d}}
\hline
\sidehead{~~\textit{Fit Parameters}}
~~~P$_{d}$ & 12.5200$\pm$0.0002 &   days & Orbital Period \\
~~~$\sqrt{e} \cos\omega_{d} $ & -0.100$\pm$0.007  &   ... & Eccentricity Reparametrization \\
~~~$\sqrt{e} \sin\omega_{d} $ & 0.093$\pm$0.008 &   ... & Eccentricity Reparametrization \\
~~~ M$_{d}$ & 139.7$^{+3.1}_{-2.8}$ &   degrees & Mean Anomaly \\
~~~i$_{d}$ & 89.4$\pm$0.3 &   degrees  & Inclination \\
~~~m$_{p,d}$ & 9.4$\pm$1.2 & M$_{\oplus}$ & Planet Mass \\
\sidehead{~~\textit{Derived Parameters}}
~~~ $\rho_{d}$ & 0.41$\pm$0.05 &  g/cc & Bulk Density \\
~~~ e$_{d}$ & 0.019$\pm$0.002 &  ... & Eccentricity \\
~~~ $\omega_{d}$ & -43$\pm$3 &  Degrees & Argument of Periastron \\
~~~ a$_{d}$ & 0.1062$\pm$0.0004 &  AU & Semi-Major Axis \\
~~~ T$_{eq,d}$ & 843$\pm$5 &  K & Equilibrium Temperature \\
\sidehead{\textbf{Planet e}}
\hline
\sidehead{~~\textit{Fit Parameters}}
~~~P$_{e}$ & 18.8009$^{+0.0006}_{-0.0005}$ &  days & Orbital Period \\
~~~$\sqrt{e} \cos\omega_{e} $ & 0.0825 $\pm$ 0.006  &  ... & Eccentricity Reparametrization \\
~~~$\sqrt{e} \sin\omega_{e} $ & -0.184 $\pm$ 0.007 &  ... & Eccentricity Reparametrization \\
~~~ M$_{e}$ & 173.7$^{+1.6}_{-1.5}$ &  degrees & Mean Anomaly \\
~~~i$_{e}$ & 89.3$\pm$0.3 &  degrees  & Inclination \\
~~~m$_{p,e}$ & 6.8$\pm$0.7 &  M$_{\oplus}$ & Planet Mass \\
\sidehead{~~\textit{Derived Parameters}}
~~~ $\rho_{e}$ & 1.41$\pm$0.19 &  g/cc & Bulk Density \\
~~~ e$_{e}$ & 0.041$\pm$0.003 &  ... & Eccentricity \\
~~~ $\omega_{e}$ & -63.3$\pm$1.2 & Degrees & Argument of Periastron \\
~~~ a$_{e}$ & 0.139$\pm$0.001 &  AU & Semi-Major Axis \\
~~~ T$_{eq,e}$ & 736$\pm$5 &  K & Equilibrium Temperature \\
\sidehead{\textbf{Planet f}}
\hline
\sidehead{~~\textit{Fit Parameters}}
~~~P$_{f}$ & 26.321 $\pm$ 0.0008 &  days & Orbital Period \\
~~~$\sqrt{e} \cos\omega_{f} $ & 0.014$^{+0.010}_{-0.006}$ &  ... & Eccentricity Reparametrization \\
~~~$\sqrt{e} \sin\omega_{f} $ & -0.013$^{+0.006}_{-0.009}$ &  ... & Eccentricity Reparametrization \\
~~~ M$_{f}$ & 51.8$^{+5.5}_{-6.1}$ &  degrees & Mean Anomaly \\
~~~i$_{f}$ & 89.4 $\pm$ 0.2 &   degrees  & Inclination \\
~~~m$_{p,f}$ & 10.8$^{+2.6}_{-2.4}$ &  M$_{\oplus}$ & Planet Mass \\
\sidehead{~~\textit{Derived Parameters}}
~~~ $\rho_{f}$ & 0.89$\pm$0.22 &  g/cc & Bulk Density \\
~~~ e$_{f}$ & 0.000$\pm$0.001 &  ... & Eccentricity \\
~~~ $\omega_{f}$ & -42.9$\pm$30 &  Degrees & Argument of Periastron \\
~~~ a$_{f}$ & 0.174$\pm$0.001 &  AU & Semi-Major Axis \\
~~~ T$_{eq,f}$ & -658$\pm$4 &  K & Equilibrium Temperature \\
\sidehead{\textbf{Planet g}}
\hline
\sidehead{~~\textit{Fit Parameters}}
~~~P$_{g}$ & 39.546 $\pm$ 0.001 &  days & Orbital Period \\
~~~$\sqrt{e} \cos\omega_{g} $ &  0.0293$^{+0.006}_{-0.005}$ &  ... & Eccentricity Reparametrization \\
~~~$\sqrt{e} \sin\omega_{g} $ & -0.18$\pm$0.01 &  ... & Eccentricity Reparametrization \\
~~~ M$_{g}$ & -120.5$^{+1.9}_{-2.1}$ &  degrees &  Mean Anomaly \\
~~~i$_{g}$ & 89.6 $\pm$ 0.2 &  degrees  & Inclination \\
~~~m$_{p,g}$ & 3.5$^{+2.8}_{-2.1}$ &  M$_{\oplus}$ & Planet Mass \\
\sidehead{~~\textit{Derived Parameters}}
~~~ $\rho_{g}$ & 3.77$\pm$1.19 & g/cc & Bulk Density \\
~~~ e$_{g}$ & 0.033$\pm$0.004 & ... & Eccentricity \\
~~~ $\omega_{g}$ & -81$\pm$2 &  Degrees & Argument of Periastron \\
~~~ a$_{g}$ & 0.229$\pm$0.001 &  AU & Semi-Major Axis \\
~~~ T$_{eq,g}$ & 574$\pm$4 &  K & Equilibrium Temperature \\
\sidehead{\textbf{TOI-1136(h)}}
\enddata
\tablenotetext{**}{ Estimated using an albedo of 0.}
\tablenotetext{$\dag$}{All of the orbital parameters presented in this table are osculating elements computed at BJD 2458680 days.}
\end{deluxetable*}

\end{document}